\documentclass[
  twocolumn,
  prb,
  showpacs,
  amsmath,
  amssymb,
  superscriptaddress
]{revtex4}

\usepackage{bm}
\usepackage{bbm}
\usepackage{graphicx}

\begin{document}

\title{Metal-insulator transition in 2D
  random fermion systems of chiral symmetry classes}

\author{E.\ J.\ K\"onig}
\affiliation{
 Inst. f\"ur Theorie der kondensierten Materie,
 Karlsruhe Institute of Technology, 76128 Karlsruhe, Germany
}
\affiliation{
 DFG Center for Functional Nanostructures,
 Karlsruhe Institute of Technology, 76128 Karlsruhe, Germany
}

\author{P.\ M.\ Ostrovsky}
\affiliation{
 Institut f\"ur Nanotechnologie, Karlsruhe Institute of Technology,
 76021 Karlsruhe, Germany
}
\affiliation{
 L.\ D.\ Landau Institute for Theoretical Physics RAS,
 119334 Moscow, Russia
}

\author{I.\ V.\ Protopopov}
\affiliation{
 Institut f\"ur Nanotechnologie, Karlsruhe Institute of Technology,
 76021 Karlsruhe, Germany
}
\affiliation{
 L.\ D.\ Landau Institute for Theoretical Physics RAS,
 119334 Moscow, Russia
}

\author{A.\ D.\ Mirlin}
\affiliation{
 Institut f\"ur Nanotechnologie, Karlsruhe Institute of Technology,
 76021 Karlsruhe, Germany
}
\affiliation{
 Inst. f\"ur Theorie der kondensierten Materie,
 Karlsruhe Institute of Technology, 76128 Karlsruhe, Germany
}
\affiliation{
 Petersburg Nuclear Physics Institute,
 188300 St.~Petersburg, Russia.
}
\affiliation{
 DFG Center for Functional Nanostructures,
 Karlsruhe Institute of Technology, 76128 Karlsruhe, Germany
}

\begin{abstract}
Field-theoretical approach to Anderson localization in 2D disordered fermionic
systems of chiral symmetry classes (BDI, AIII, CII) is developed. Important
representatives of these symmetry classes are random hopping models on bipartite
lattices at the band center. As was found by Gade and Wegner two decades ago
within the sigma-model formalism, quantum interference effects in these classes
are absent to all orders of perturbation theory. We demonstrate that the quantum
localization effects emerge when the theory is treated non-perturbatively.
Specifically, they are controlled by topological vortex-like excitations of the
sigma models. We derive renormalization-group equations including these
non-perturbative contributions. Analyzing them, we find that the 2D disordered
systems of chiral classes undergo a metal-insulator transition driven by
topologically induced Anderson localization. We also show that the Wess-Zumino
and $\mathbb{Z}_2$ theta terms on surfaces of 3D topological insulators (in
classes AIII and CII, respectively) overpower the vortex-induced localization.

\end{abstract}

\pacs{%
73.23.-b, 
73.20.Fz, 
73.43.Nq  
}

\maketitle

\section{Introduction}

Disorder-induced metal-insulator transitions represent a fundamental concept of
modern condensed matter physics. The principal mechanism of this phenomenon is
Anderson localization \cite{Anderson58} due to quantum interference between
different trajectories of electron propagation in a disordered medium. As was
realized in seventies by Thouless and Wegner \cite{Thouless74,Wegner76}, the
problem of Anderson localization is closely connected to the scaling theory of
critical phenomena. Several years later, Abrahams and coauthors formulated the
scaling theory of localization \cite{AALR} describing the problem in terms of a
flow of the dimensionless conductance $g$ with the system size $L$. This
phenomenological theory was put on a solid basis by Wegner \cite{Wegner79} who
discovered the field-theoretical description of Anderson localization in terms
of a nonlinear sigma model. The latter is an effective field theory of the
problem, capturing all essential properties of the system including its symmetry
and topology. The crucial importance of the topological aspect was first
demonstrated by Pruisken in the framework of the quantum Hall effect
\cite{Pruisken84}. The field-theoretical (sigma model) formulation of the
problem underpins the notion of universality of critical behavior at
metal-insulator transitions, i.e., its independence of microscopic details of
the electron motion. Over the years the sigma model permitted the great progress
in understanding localization phenomena including, in particular, critical
behavior,\cite{Wegner79, Efetov97, Evers08} energy level and wave function
statistics,\cite{Efetov97, Mirlin00} symmetry classification,\cite{Zirnbauer96}
interaction (most prominently 2D metal-insulator transition) and non-equilibrium
effects,\cite{Finkelstein90, BelitzKirkpatrick, KamenevAndreev, Punnoose} theory
of unconventional disordered superconductors,\cite{ASZ} classification of
topological insulators and superconductors.\cite{SchnyderKitaev}

Recent years have witnessed increasing interest and significant progress in the
field of the Anderson localization. Development of the complete symmetry
classification of disordered systems \cite{Zirnbauer96} is one of most important
advances. It has been realized that underlying symmetries and topologies induce
a remarkable variety of the types of infrared (long-distance, low-temperature)
behavior, including, in particular, critical phases and quantum phase
transitions between metallic and insulating states \cite{Evers08}. Experimental
discoveries of graphene \cite{graphene} and of topological insulators
\cite{Molenkamp, HasanKane10, QiZhang11} have given an additional boost to the
field. Disordered Dirac fermions in graphene in the absence of valley mixing and
surface states of disordered topological insulators and superconductors are
characterized by sigma models with Wess-Zumino or theta terms, ensuring a
topological protection from localization \cite{Evers08, SchnyderKitaev, AL50}. 

The full symmetry classification of disordered fermion systems
\cite{Zirnbauer96, Evers08} includes three families of symmetry classes:
conventional (Wigner-Dyson), chiral, and superconducting (Bogoliubov-de Gennes).
In this paper we will consider localization properties of two-dimensional (2D) 
chiral models. Characteristic feature of these systems  is the chiral symmetry
of the underlying Hamiltonian $H = -C H C^{-1}$, where  $C^2 = 1$. It is
convenient to use the block structure such that $C$ is represented by the third
Pauli matrix $\tau_z$; then the Hamiltonian acquires a block off-diagonal form,
\begin{equation}
 H
  = \begin{pmatrix}
      0 & h \\
      h^\dagger & 0
    \end{pmatrix}.
 \label{H-chiral}
\end{equation}

One-dimensional disordered systems with chiral symmetry were studied in Refs.\
\onlinecite{chiral1D}. An effective field-theoretical description -- non-linear
sigma model -- for chiral Hamiltonians with randomness was developed in Refs.\
\onlinecite{chiralsigmamodel, GadeWegner}. Spectral properties of chiral random
matrix ensembles were discussed in the context of mesoscopic transport and
quantum chromodynamics in Refs.\ \onlinecite{chiralRMT, Ivanov02}. 

A standard realization of a chiral system is provided by a bipartite lattice
model (i.e., containing two sublattices) with random hopping between the
sublattices and no diagonal disorder. The Hamiltonian of such a model at the
band center is manifestly block off-diagonal in the sublattice space. In analogy
with Wigner-Dyson classes, there are three chiral classes: chiral unitary
(AIII), chiral orthogonal (BDI), and chiral symplectic (CII) classes.  The class
AIII describes chiral systems with broken time-reversal symmetry, which may be
due to magnetic flux (random or homogeneous) through the lattice plaquettes. The
classes BDI and CII are characterized, in addition to the chiral symmetry, by
time-reversal invariance, $H = K H^T K^{-1}$, where the unitary matrix $K$
satisfies $K^2 = \pm 1$. For the chiral orthogonal class BDI one has $K^2 = 1$;
in the standard representation $K = 1$. For the chiral symplectic class CII,
which corresponds to the case of broken spin rotation symmetry (usually due to
spin-orbit interaction), $K^2 = -1$; in the standard representation $K =
i\sigma_y$, where $\sigma_y$ is the second Pauli matrix in the spin space. Thus,
in the standard form, random Hamiltonians of the chiral classes have block
off-diagonal structure and consist of complex, real, and real quaternion entries
for the classes AIII, BDI, and CII, respectively.

Realizations of 2D chiral classes arise naturally when one models the disordered
graphene at the Dirac point \cite{OGM}. Specifically, the chiral symmetry occurs
if the dominant disorder is of random-hopping character. An important example 
is the random magnetic flux, either genuine or effective (due to lattice
corrugations), in which case the system belongs to the chiral unitary class
AIII. Another possible realization is provided by graphene with vacancies. A
vacancy can be modeled by cutting all lattice bonds adjacent to the vacated
site, so that vacancies represent a special type of bond disorder. This model is
in the chiral orthogonal symmetry class BDI. If an external magnetic field is
applied to graphene with vacancies, the time-reversal symmetry is broken and the
system falls into the class AIII. Taking into account the spin-orbit interaction
(in the absence of magnetic field) brings the system into the chiral symplectic
symmetry class CII. 

Clearly, the above symmetry analysis is not specific for graphene (hexagonal
lattice) but is equally applicable to a random-hopping model on any other
bipartite lattice (e.g. the square lattice) at the band center. The analysis of
Anderson localization in the chiral classes performed in the present work is
equally applicable to all such models.  

A remarkable property of the three chiral symmetry classes in 2D is the exact
absence of weak localization corrections to all orders in the perturbation
theory \cite{GadeWegner}. At the same time, when the disorder is strong enough,
the transition to insulating state is inevitable. Indeed, consider a
tight-binding Hamiltonian on a 2D bipartite (e.g., square or honeycomb) lattice.
Let us introduce disorder by randomly cutting the lattice bonds. (As discussed
above, this model belongs to the chiral orthogonal class BDI.)  Classically,
this is the standard percolation model with bond disorder. When the
concentration of removed bonds reaches the critical value ($1/2$ for square
lattice and $1 - 2\sin(\pi/18) \approx 0.65$ for honeycomb lattice), the system
becomes disconnected and undergoes the classical percolation transition to the
insulating state. Obviously, the conductivity of such a system (defined in the
limit of the infinite system size) is zero. It is well known that, in
Wigner-Dyson symmetry classes, such a classical percolation transition is
preceded by the Anderson metal-insulator transition. Specifically, due to
quantum interference, the conductivity of a quantum system becomes zero at the
point where classically it would still conduct (the remaining bonds are still
percolating). Does this happen also for models of chiral classes? The exact
absence of quantum-interference effects to all orders \cite{GadeWegner} suggests
that the answer is negative. However, numerical simulations in Refs.\
\onlinecite{Motrunich02, Bocquet03} do show a quantum localization transition in
chiral classes. These papers explored numerically models with continuous
disorder which are always conducting classically, and yet found a
metal-insulator transition. Thus, there appears to be a contradiction between
the absence of localization in the sigma-model analysis of Ref.\
\onlinecite{GadeWegner} and the numerical results. This contradiction is
resolved in the present paper.

We will show below that, while the analysis of Ref.\ \onlinecite{GadeWegner}
is fully correct to all orders of perturbation theory, the quantum interference
effects do emerge in the sigma-model field theory when one considers
non-perturbative effects related to topologically non-trivial, vortex-like
excitations. The field-theoretical analysis of the quantum localization in
chiral classes requiring a systematic study of such non-perturbative effects
constitutes the subject of the paper.

Non-perturbative effects caused by vortices were considered in the works
of Berezinskii\cite{Berezinskii71} and Kosterlitz and
Thouless\cite{KosterlitzThouless} in the context of planar 2D ferromagnet (XY
model) and 2D superfluid flow. These systems are characterized by a
$\mathrm{U}(1)$ field (direction of magnetization and phase of the condensate
wa)ve function, respectively) and exhibit the temperature-driven
Berezinskii-Kosterlitz-Thouless (BKT) phase transition between quasi-ordered
(ferromagnetic or superfluid) and disordered (paramagnetic or normal fluid)
phases caused by vortex unbinding. Renormalization group (RG) analysis of this
phase transition was first developed by Kosterlitz\cite{Kosterlitz74}.
Alternative descriptions of the same systems are provided by the Coulomb gas
(see Ref.\ \onlinecite{Minnhagen} for a review) and sine-Gordon models.
RG equations for the sine-Gordon model, equivalent to Kosterlitz RG, were
derived in Ref.\ \onlinecite{Wiegman} in the one-loop order and later in Ref.\
\onlinecite{Amit} with the two-loop accuracy. We will show that, in analogy
with the BKT physics, vortices drive the localization transition in chiral
classes. However, the emerging theory is more complicated and the character of
transition is essentially different, since the symmetry group of the
localization problem is different from $\mathrm{U}(1)$.

The layout of the paper is as follows. In Sec.\ \ref{sec:pre}, we present the
general structure of the sigma model for chiral symmetry classes, outline the
proof of the absence of perturbative weak localization corrections, and identify
the mechanism of non-perturbative effects. The general background-field
formalism for the renormalization of the sigma model is developed in Sec.\
\ref{sec:background} and then applied to the calculation of perturbative
corrections in Sec.\ \ref{sec:perturb}.  In Sec.\ \ref{sec:nonperturb}, we
construct the vortex excitations and explore their effect on the conductivity.
For this purpose, we derive a set of RG equations on the conductivity $\sigma$,
Gade-term coupling $c$ (controlling the behavior of the density of states), and
the vortex fugacity $y$. Its analysis reveals a metal-insulator transition.
Therefore, we show that the chiral classes do exhibit quantum interference
effects -- the weak localization and the Anderson transition -- driven by
topological excitations in the sigma model. In Sec.\ \ref{sec:sine-gordon} we
develop an alternative approach to non-perturbative renormalization in class
AIII. This approach is based on a mapping of the original sigma-model with
vortices onto a dual theory which has a character of generalized sine-Gordon
theory. Section \ref{sec:topology} is devoted to the interplay of this
topological localization with further possible topological properties of 2D
chiral-class systems: the Wess-Zumino term in class AIII and the $\mathbb{Z}_2$
theta term in class CII. These terms in the sigma models emerge when the 2D
systems are realized on surfaces of 3D topological insulators of the
corresponding symmetry classes. We show that the vortex-induced topological
localization becomes inefficient in the presence of such terms. The latter thus
ensure the topological protection of the system from localization. Our results
are summarized in Sec.\ \ref{sec:summary}. Some technical details are presented
in two Appendices.

\section{Preliminaries}
\label{sec:pre}

Very generally, localization properties of a disordered system are described by
the non-linear sigma model field theory. We will use the fermionic replica
version of the sigma model. (Alternatively, the same analysis can be performed
within the supersymmetry approach; we choose the replica formalism here since in
this case the presentation turns out to be physically more transparent.) The
sigma-model target space is a compact symmetric space, see Ref.\
\onlinecite{Evers08} for a review. For the three chiral classes, the field
$Q(\mathbf{r})$ is a unitary matrix taking values in the symmetric spaces listed
in Table \ref{tab:classes}. The size of the $Q$ matrix, $N$, is determined by
the number of replicas; in the end one should take the replica limit $N\to 0$.
In each of the classes CII and BDI, the $Q$ matrix obeys a certain linear
constraint. These constraints can be represented as $Q = \bar Q$, where the
``bar'' operation is defined in the following way:
\begin{equation}
 \bar Q
  = \begin{cases}
      Q^T, & \text{CII}, \\
      C Q^T C, & \text{BDI}.
    \end{cases}
 \label{bar}
\end{equation}
In the class BDI, $C$ is a skew-symmetric matrix such that $C^2 = 1$. In the
standard parametrization of the model, the number of replicas $N$ is even and
includes the spin degree of freedom while the matrix $C$ is the $\sigma_y$ Pauli
matrix acting in the spin subspace.

\begin{table}
\tabcolsep=4pt
\begin{tabular}{ccccccc}
\hline\hline
Class  & Manifold & $Q$ matrix & $D$ & $s$\\
\hline
AIII
  & $\mathrm{U}(N)$
  & No restriction
  & $N^2$
  & 1 \\
CII
  & $\mathrm{U}(N) / \mathrm{O}(N)$
  & $Q = Q^T$
  & $N(N+1)/2$
  & 2 \\
BDI
  & $\mathrm{U}(N) / \mathrm{Sp}(N)$
  & $Q = C Q^T C$
  & $N(N-1)/2$
  & 2 \\
\hline\hline
\end{tabular}
\caption{List of chiral symmetry classes including corresponding compact
sigma-model manifolds, symmetry constraints on the $Q$ matrix, and the
dimensionalities $D$ of the target spaces. Parameter $s$ in the last column is
introduced to normalize the conductivity in Eq.\ (\protect\ref{action}).}
\label{tab:classes}
\end{table}

The distinguishing feature of the three manifolds of chiral-class sigma models
is that the corresponding tangent spaces possess a unit generator commuting with
all other (traceless) generators. This generator corresponds to the
$\mathrm{U}(1)$ part of the corresponding symmetric space. The presence of this
special generator allows an additional Gade term in the sigma model action
\cite{GadeWegner}:
\begin{equation}
 S[Q]
  = -\int \frac{d^2r}{8\pi s} \left[
      \sigma \mathop{\mathrm{Tr}}(Q^{-1} \nabla Q)^2
      +c (\mathop{\mathrm{Tr}} Q^{-1} \nabla Q)^2
    \right].
 \label{action}
\end{equation}
The parameter $\sigma$ is the dimensionless conductivity of the sample
(measured in units $e^2/ \pi h$)
and $c$ multiplies the Gade term. The bare value of $c$ is of order $1$.

Let us decompose $Q$ according to $Q = e^{i\phi} U$ with $\det U = 1$. The
overall phase $\phi$ is exactly the $\mathrm{U}(1)$ degree of freedom
characteristic for sigma models of the chiral symmetry classes. The action
(\ref{action}) takes the form
\begin{equation}
 S[Q]
  = \int \frac{d^2r}{8\pi s} \left[
      N (\sigma + N c) (\nabla \phi)^2
      +\sigma \mathop{\mathrm{Tr}}(\nabla U^{-1} \nabla U)
    \right].
 \label{Gade}
\end{equation}
We see that variables $\phi$ and $U$ decouple and that the action is quadratic
in $\phi$. This means that the prefactor $\sigma + N c$ of the $(\nabla \phi)^2$
term is not renormalized. In the replica limit $N \to 0$, the absence of
localization corrections to $\sigma$ follows, which is the result of Gade and
Wegner\cite{GadeWegner}.

The above argument for the absence of the renormalization of conductivity has,
however, a caveat. Specifically, the action for the variable $\phi$ is not
strictly Gaussian in view of the compact $\mathrm{U}(1)$ nature of this degree
of freedom. The group $\mathrm{U}(1)$ is not simply connected allowing for
topological excitations -- vortices. Fluctuations of the matrix $Q$ involving
vortices yield corrections to $\sigma + N c$ via the BKT\cite{Berezinskii71,
KosterlitzThouless} mechanism. These corrections break the replica symmetry and
thus invalidate the Gade and Wegner argument. Our goal is to analyze the effect
of vortices on the renormalization of the chiral-class sigma models. We will
start with the general background field formulation of the renormalization. Then
we will consider both perturbative and non-perturbing (due to vortices)
contributions to the RG equations.

\section{Background field formalism}
\label{sec:background}

Very generally, renormalization of the sigma model can be implemented in the
framework of the background field formalism \cite{Polyakov75, Pruisken87}. 
Following this approach, we first separate slow and fast parts in the matrix
$Q$. The former is referred to as the background field. Then the partition
function is integrated over the fast part of $Q$ yielding the effective action
for the background field. This effective action has the same sigma-model form
with the renormalized parameters $\sigma$ and $c$.

In order to renormalize the action (\ref{action}), we parametrize $Q = \bar V
\tilde Q V$, where $\tilde Q$ is the fast field of the same symmetry as $Q$ and
unitary matrices $V$ and $\bar V$ represent the slow background field. In the
symmetry class AIII, matrices $V$ and $\bar V$ are independent while for the
other two chiral classes they are related by Eq.\ (\ref{bar}). After integrating
out $\tilde Q$, the effective action involves only the gauge invariant
combination of slow fields $Q' = \bar V V$. It is convenient to introduce the
following notations for the gradients of the slow field:
\begin{equation}
 \mathbf{A}
  = \nabla V V^{-1},
 \qquad
 \bar{\mathbf{A}}
  = \bar V^{-1} \nabla \bar V.
 \label{AbarA}
\end{equation}
Matrices $\mathbf{A}$ and $\bar{\mathbf{A}}$ are related by Eq.\ (\ref{bar}) in
classes CII and BDI.

We start the renormalization group analysis with the action $S_0[Q]$ given by
Eq.\ (\ref{action}) with bare parameters $\sigma_0$ and $c_0$. Upon substitution
$Q = \bar V \tilde Q V$, the action decomposes into four terms:
\begin{equation}
 S_0[Q]
  = S_0[Q'] + S_L[\tilde Q]
    +S_\mathrm{int}^{(1)}[\bar V, V, \tilde Q]
    +S_\mathrm{int}^{(2)}[\bar V, V, \tilde Q].
 \label{varsep}
\end{equation}
The first term, $S_0[Q']$, is the bare action for the slow background field $Q'
= \bar V V$. It can be represented as
\begin{equation}
 S_0[Q']
  = -\!\!\int\! \frac{d^2r}{8\pi s} \left\{\!
      \sigma_0 \mathop{\mathrm{Tr}} \left(
          \bar{\mathbf{A}} + \mathbf{A}
      \right)^2
      +c_0 \left[ \mathop{\mathrm{Tr}} \left(
          \bar{\mathbf{A}} + \mathbf{A}
      \right) \right]^2
    \right\}.
 \label{S0}
\end{equation}
The action for the fast field $S_L[\tilde Q]$ is given by the same Eq.\
(\ref{action}) with bare parameters $\sigma_0$ and $c_0$ but includes also the
mass term,
\begin{equation}
 S_L[\tilde Q]
  = S_0[\tilde Q] - \frac{\sigma_0}{8\pi s L^2} \int d^2r \mathop{\mathrm{Tr}}
    \big( 2 - \tilde Q - \tilde Q^{-1} \big).
 \label{SL}
\end{equation}
The mass $\sim 1/L$ is introduced to ensure that the matrix $\tilde Q$ contains
only fast degrees of freedom. At the same time, we will assume that the slow
fields $V$ and $\bar V$ change very little on the distances of the order of $L$.
It is important that the mass term is symmetric on the manifold of $\tilde Q$;
this will guarantee that the renormalized action involves only the
gauge-invariant combination $\bar V V$ and retains the full symmetry of the
sigma model. Finally, the terms in the action representing the interaction
between the slow and fast modes read
\begin{align}
 S_\mathrm{int}^{(1)}
  &=-\int \frac{d^2r}{4\pi s} \Big[
      \sigma_0 \mathop{\mathrm{Tr}} \left(
        \nabla \tilde Q \tilde Q^{-1} \bar{\mathbf{A}}
        +\tilde Q^{-1} \nabla \tilde Q \mathbf{A}
      \right) \notag \\
      &\hspace{15mm}+c_0 \mathop{\mathrm{Tr}} \tilde Q^{-1} \nabla \tilde Q
        \mathop{\mathrm{Tr}} \left(
          \bar{\mathbf{A}} + \mathbf{A}
        \right)
    \Big], \\
 S_\mathrm{int}^{(2)}
  &=\sigma_0 \int \frac{d^2r}{4\pi s} \mathop{\mathrm{Tr}} \left[
        \bar{\mathbf{A}} \mathbf{A}
        -\tilde Q^{-1} \bar{\mathbf{A}} \tilde Q \mathbf{A}
    \right].
 \label{Sint}
\end{align}

Integrating $e^{-S[Q]}$ over fast modes $\tilde Q$, taking the logarithm of the
result, performing the gradient expansion, and retaining the terms with two
spatial derivatives (i.e. up to second order in $\mathbf{A}$ and
$\bar{\mathbf{A}}$), we obtain the effective action for the slow field:
\begin{equation}
 S[Q']
  = S_0[Q'] + \left< S_\mathrm{int}^{(2)} \right>
    -\frac{1}{2} \left<
      \big[S_\mathrm{int}^{(1)}\big]^2
    \right>.
 \label{renorm}
\end{equation}
Here $\langle \dots \rangle$ denotes the averaging over $\tilde Q$ with the
weight $e^{-S_L[\tilde Q]}$. Note that the average $\langle S_\mathrm{int}^{(1)}
\rangle$ is linear in the gradients of the slow variables and is hence zero. The
renormalizability of the sigma model implies that the effective action for the
slow field $Q'$ has the form of Eq.\ (\ref{S0}) with the bare parameters
$\sigma_0$ and $c_0$ replaced with their renormalized values $\sigma$ and $c$.
Using this fact, we will assume three particular forms of the background field
and thus establish the renormalized parameters in terms of $\tilde Q$ integrals
for all three symmetry classes.

The $D$-dimensional sigma-model target space is generated by $D - 1$ traceless
Hermitian generators and one generator proportional to the unit matrix. In the
classes CII and BDI these generators are subject to an additional symmetry
constraint $T = \bar T$. We normalize the generators as $\mathop{\mathrm{Tr}}
T^a T^b = \delta^{ab}$. The traceless generators of the sigma-model manifold
obey the following Fierz identities:
\begin{equation}
 \sum_a T^a_{ij} T^a_{kl}
  = \begin{cases}
      \delta_{il} \delta_{jk} - \dfrac{1}{N}\, \delta_{ij} \delta_{kl},
        & \text{AIII}, \\[6pt]
      \dfrac{1}{2} \left(
        \delta_{il} \delta_{jk} + \delta_{ik} \delta_{jl}
      \right) - \dfrac{1}{N}\, \delta_{ij} \delta_{kl},
        & \text{CII}, \\[6pt]
      \dfrac{1}{2} \left(
        \delta_{il} \delta_{jk} + C_{ik} C_{jl}
      \right) - \dfrac{1}{N}\, \delta_{ij} \delta_{kl},
        & \text{BDI}.
    \end{cases}
 \label{FierzT}
\end{equation}

Assume the background field of the form $\mathbf{A} = \bar{\mathbf{A}} = i
\mathbf{J} T$ with some constant vector $\mathbf{J}$ and $T$ being one of the $D
- 1$ traceless generators of the sigma-model manifold. Substituting this
background field into Eq.\ (\ref{renorm}), averaging over directions of
$\mathbf{J}$, and comparing the prefactors of $\mathbf{J}^2$ in the left- and
right-hand side of the equation, we find the renormalized conductivity in the
form
\begin{multline}
 \sigma
  = \sigma_0
    -\frac{\sigma_0}{4 A} \int d^2r \mathop{\mathrm{Tr}} \left<
      \big[ \tilde Q, T \big] \big[ \tilde Q^{-1}, T \big]
    \right> \\
    +\frac{\sigma_0^2}{32\pi s A} \left< \left[
      \int d^2r \mathop{\mathrm{Tr}}
      \big\{\nabla \tilde Q, \tilde Q^{-1} \big\} T
    \right]^2 \right>.
 \label{bf1}
\end{multline}

Next, we assume a pure gauge background field such that $\mathbf{A} =
-\bar{\mathbf{A}} = i \mathbf{J} t$. The matrix $t$ is now one of generators of
the stabilizer group, $\mathrm{U}(N)$, $\mathrm{O}(N)$, and $\mathrm{Sp}(N)$,
for the classes AIII, CII, and BDI, respectively. These generators obey the
second Fierz identity
\begin{equation}
 \sum_a t^a_{ij} t^a_{kl}
  = \begin{cases}
      \delta_{il} \delta_{jk},
        & \text{AIII}, \\[6pt]
      \dfrac{1}{2} \left(
        \delta_{il} \delta_{jk} - \delta_{ik} \delta_{jl}
      \right),
        & \text{CII}, \\[6pt]
      \dfrac{1}{2} \left(
        \delta_{il} \delta_{jk} - C_{ik} C_{jl}
      \right),
        & \text{BDI}.
    \end{cases}
 \label{Fierzt}
\end{equation}
Such a gauge background field corresponds to the constant $Q' = 1$ and hence the
left-hand side of Eq.\ (\ref{renorm}) vanishes. Averaging over directions of
$\mathbf{J}$, we obtain the following Ward identity
\begin{multline}
 0
  = \frac{\sigma_0}{4 A} \int d^2r \mathop{\mathrm{Tr}} \left<
      \big[ \tilde Q, t \big] \big[ \tilde Q^{-1}, t \big]
    \right> \\
    +\frac{\sigma_0^2}{32\pi s A} \left< \left[
      \int d^2r \mathop{\mathrm{Tr}}
      \big[ \nabla \tilde Q, \tilde Q^{-1} \big] t
    \right]^2 \right>.
 \label{bf2}
\end{multline}

We sum up identities (\ref{bf1}) with all generators $T$ and add identities
(\ref{bf2}) with all possible matrices $t$. With the help of the Fierz
identities (\ref{FierzT}) and (\ref{Fierzt}) and using the properties $Q = \bar
Q$ in the classes CII and BDI, we obtain the renormalized conductivity
\begin{equation}
 \sigma
  = \sigma_0 + \frac{\sigma_0^2}{D - 1} \left(
      B_1 - \frac{B_2}{N}
    \right).
 \label{sigma}
\end{equation}
Here we have introduced the averages
\begin{align}
 B_1
  &=\int\frac{d(\mathbf{r - r'})}{8 \pi s} \left<
      \mathop{\mathrm{Tr}} \big( \tilde Q^{-1} \nabla \tilde Q \big)_r
      \big( \tilde Q^{-1} \nabla \tilde Q \big)_{r'}
    \right>, \label{B1} \\
 B_2
  &=\int\frac{d(\mathbf{r - r'})}{8 \pi s} \left<
      \mathop{\mathrm{Tr}} \big( \tilde Q^{-1} \nabla \tilde Q \big)_r
      \mathop{\mathrm{Tr}} \big( \tilde Q^{-1} \nabla \tilde Q \big)_{r'}
    \right>. \label{B2}
\end{align}

Finally, consider the background field of the form $\mathbf{A} =
\bar{\mathbf{A}} = i \mathbf{J} \mathbbm{1}/\sqrt{N}$ generated by the unit
matrix. This field configuration brings into play both the kinetic and the Gade
terms of the action. After averaging over directions of $\mathbf{J}$, we obtain
the identity
\begin{equation}
 \sigma + N c
  = \sigma_0 + N c_0
    +\frac{(\sigma_0 + N c_0)^2}{N}\; B_2.
 \label{bf3}
\end{equation}
With $\sigma$ from Eq.\ (\ref{sigma}), this yields the following expression for
the renormalized coupling $c$:
\begin{equation}
 c
  = c_0 - \frac{\sigma_0^2\, B_1}{N (D - 1)} + \left[
      \frac{\sigma_0^2}{D - 1} + (\sigma_0 + N c_0)^2
    \right] \frac{B_2}{N^2}.
 \label{c}
\end{equation}

Equations (\ref{sigma}) and (\ref{c}) represent renormalization of $\sigma$
and $c$ in the most general form. The averages $B_{1,2}$ implicitly depend on
the infrared scale $L$ set by the mass term in the fast field action (\ref{SL}).
The RG flow equations are obtained in the standard way by taking the derivative
of Eqs.\ (\ref{sigma}) and (\ref{c}) with respect to $\ln L$ and replacing
$\sigma_0$ and $c_0$ in the right-hand side of these equations with the running,
$L$-dependent values $\sigma$ and $c$ of the couplings.

\section{Perturbative renormalization}
\label{sec:perturb}

In the previous section we have developed the background field formalism and
reduced the renormalization of the sigma models of chiral classes to the
calculation of the correlators (\ref{B1}) and (\ref{B2}). The matrix $\tilde Q$
belongs to the corresponding sigma-model manifold and averaging is performed
with the statistical weight determined by the action (\ref{SL}). We will now
perform this averaging using a saddle point method. The saddle point
approximation is justified in the limit $\sigma_0 \gg 1$. We start with the
spatially uniform saddle point $\tilde Q = 1$. Note that the mass term in the
action (\ref{SL}) is minimized by this, rather than any other, constant value of
$\tilde Q$. Expansion in the vicinity of the spatially uniform saddle point
yields perturbative contributions to the renormalized couplings. As we are going
to explain, such perturbative correction to $\sigma$ vanishes, in the replica
limit $N \to 0$, in all orders of the perturbation theory for all three chiral
classes. We will then proceed by including a non-perturbative contribution from
saddle configurations containing a vortex-antivortex dipole. We will show that
this yields a non-zero result. 

We parametrize the fast field $\tilde Q = 1 + i W - W^2/2 + O(W^3)$ by a
Hermitian matrix $W$ subjected to the linear constraint $W = \bar W$ in classes
CII and BDI. Let us decompose $W$ in generators of the corresponding symmetric
space: $W = w_0 \mathbbm{1}/\sqrt{N} + \sum_a w_a T^a$. Expanding Eq.\
(\ref{SL}) to the second order in $W$, we obtain the Gaussian action
\begin{multline}
 S_L[W]
  = \int \frac{d^2r}{8\pi s} \bigg\{
      (\sigma_0 + N c_0) (\nabla w_0)^2 + \sigma_0 L^{-2} w_0^2 \\
      +\sigma_0 \sum_a \Big[
        (\nabla w_a)^2 + L^{-2} w_a^2
      \Big]
    \bigg\}.
 \label{SLGauss}
\end{multline}
This action yields the following propagators of the components of $W$:
\begin{align}
 \langle w_0(\mathbf{q}) w_0(-\mathbf{q}) \rangle
  &=\frac{4\pi s}{(\sigma_0 + N c_0) q^2 + \sigma_0 L^{-2}}, \label{w0w0} \\
 \langle w_a(\mathbf{q}) w_a(-\mathbf{q}) \rangle
  &=\frac{4\pi s}{\sigma_0 (q^2 + L^{-2})}. \label{wawa}
\end{align}

Within the perturbative calculation, the average $B_2$ is identically zero since
it involves only the $\mathrm{U}(1)$ component of $\tilde Q$; specifically,
$\mathop{\mathrm{Tr}} \tilde Q^{-1} \nabla \tilde Q = i\sqrt{N} \nabla w_0$.
Consequently, $B_2$ vanishes, $B_2 \propto q^2 \langle w_0 (\mathbf{q})
w_0(-\mathbf{q}) \rangle_{q\to 0} = 0$, that implies that $\sigma + Nc$ is not
renormalized, see Eq.\ (\ref{bf3}). This cancellation of $B_2$ is the essence of
the Gade-Wegner argument \cite{GadeWegner} for the absence of weak localization
corrections to conductivity. We will see later that the saddle points involving
vortices yield a nonvanishing contribution to $B_2$.

Contrary to $B_2$, the average $B_1$ is non-zero already on the perturbative
level. We will calculate it within the one-loop approximation. Rewriting Eq.\
(\ref{B1}) in the momentum representation, expanding in the components of $W$,
and applying the Wick theorem, we find
\begin{multline}
 B_1
  = \int \frac{q^2\, d^2q}{32 \pi^3 s} \Big[
      \big< W_{ij}(\mathbf{q}) W_{kl}(-\mathbf{q}) \big>
      \big< W_{jk}(\mathbf{q}) W_{li}(-\mathbf{q}) \big> \\
      -\big< W_{ij}(\mathbf{q}) W_{jk}(-\mathbf{q}) \big>
      \big< W_{kl}(\mathbf{q}) W_{li}(-\mathbf{q}) \big>
    \Big].
\end{multline}
We have dropped the term $\sim q^2 \langle W^2 \rangle|_{q \to 0}$. The absence
of such a term is justified by the finite mass in the action (\ref{SLGauss}) and
hence a finite limit of the $W$ propagator at $q \to 0$. Using the decomposition
of $W$ in generators $T$ and the propagators (\ref{w0w0}) and (\ref{wawa}), we
obtain $B_1$ in the form
\begin{equation}
 B_1
  = \int^\Lambda \frac{s\; q^3\; dq}{2 \sigma_0^2 (q^2 + L^{-2})^2}
    \sum_{a,b} \mathop{\mathrm{Tr}} \big[ T^a, T^b \big]^2.
 \label{B1_1loop}
\end{equation}
Note that the propagator $\langle w_0 w_0 \rangle$ is canceled in the expression
for $B_1$. The momentum integral is logarithmically divergent and we have
introduced the ultraviolet cut-off $\Lambda$ to regularize this divergence.

Using the appropriate Fierz identity, we calculate the trace of products of four
generators and obtain the result
\begin{equation}
 B_1
  \simeq N (1 - D)\frac{\ln (\Lambda L)}{\sigma_0^2}.
\end{equation}
Substituting this result and $B_2 = 0$ into Eqs.\ (\ref{sigma}) and (\ref{c}),
we obtain logarithmic corrections to $\sigma$ and $c$ that can be recast in the
form of renormalization group equations
\begin{align}
 \frac{\partial \sigma}{\partial \ln L}
  &= -N + N O(1/\sigma), \label{RGsigma} \\
 \frac{\partial c}{\partial \ln L}
  &= 1 + O(1/\sigma). \label{RGc}
\end{align}
These equations describe the real space scaling with the running infrared
cut-off length $L$. The higher loop contributions provide $O(1/\sigma)$ terms in
the perturbative RG equations. In the replica limit, renormalization of $\sigma$
vanishes, in full agreement with the argument by Gade and Wegner
\cite{GadeWegner}.

A remarkable feature of the perturbative renormalization in class AIII is that
the right-hand side of Eq.\ (\ref{RGc}) is exactly $1$ in the limit $N \to 0$.
Higher loop contributions to the beta function for $c$ are proportional to $N$
and vanish in the replica limit.\cite{Guruswamy, Elio} This is not the case in
the other two chiral symmetry classes. The exact perturbative RG equations in
class AIII can be applied even when $\sigma$ is not large. We will use this fact
below for constructing the non-perturbative RG flow diagram.

\section{Non-perturbative renormalization}
\label{sec:nonperturb}

In order to find a correction to conductivity that does not vanish in the limit
$N \to 0$, we have to consider the saddle-point configurations other than
$\tilde Q = 1$. These configurations will include vortices. Vortices are
singular points of the $Q$ matrix such that the overall phase $\phi$ of $Q$
rotates by $2\pi$ along any path going around the vortex.

The sigma-model description with small gradients of $Q$ breaks down at distances
of the order of the mean free path from the vortex center. This region, referred
to as the vortex core, should be excluded from the sigma model action
(\ref{action}). The contribution of the vortex core to the overall action,
$S_v$, is not universal and depends on details of the ballistic electron
dynamics inside the core. Generically, $S_v \propto \sigma_0$. This introduces a
new coupling constant in our theory, referred to as fugacity, $y_0 = e^{-S_v}$,
that is the statistical weight associated with the vortex core.

\subsection{Vortex configurations}

Let us construct explicitly the vortex configuration in three chiral classes.
The minimal model of classes AIII and CII with a single replica reduces to the
abelian $\mathrm{U}(1)$ model with $\tilde Q = e^{i\phi}$. The same situation
occurs in class BDI at the minimum number of replicas $N = 2$. In the latter
case the matrix $\tilde Q$ retains the $2 \times 2$ spin structure: $\tilde Q =
e^{i\phi} \mathbbm{1}$. In all three models, the action (\ref{SL}) has the form
(we neglect the mass term for simplicity)
\begin{equation}
 S[\tilde Q]
  = \frac{K}{2\pi} \int d^2 r (\nabla \phi)^2,
 \label{SL_phase}
\end{equation}
where we have introduced the stiffness parameter
\begin{equation}
 K
  = \begin{cases}
      (\sigma + c)/4, & \text{AIII}, \\
      (\sigma + c)/8, & \text{CII}, \\
      (\sigma + 2 c)/4, & \text{BDI}.
    \end{cases}
 \label{K}
\end{equation}

Assume a set of vortices with positions $\mathbf{r}_i$ and charges (vorticities)
$n_i = \pm 1$. Such a configuration creates the gradient of
$\phi$ given by
\begin{equation}
 \nabla_a \phi(\mathbf{r})
  = \epsilon_{ab} \sum_i n_i \nabla_b \ln |\mathbf{r} - \mathbf{r}_i|.
 \label{current}
\end{equation}
Indeed, this field $\phi$ satisfies the equation of motion $\nabla^2 \phi = 0$
with additional condition of $2\pi n_i$ winding of $\phi$ around every vortex.
The mass term in the action (\ref{SL}), which we have temporarily neglected in
Eq.\ (\ref{SL_phase}), requires $\phi = 0$ at infinity. This is only possible if
the system of vortices is neutral: $\sum_i n_i = 0$. Thus the simplest allowed
configuration of vortices is a vortex-antivortex pair, i.e., a dipole.
Configurations with more than two vortices or vortices with higher winding
numbers have lower statistical weight and can be safely neglected.

Let us consider an individual dipole with vortex and antivortex at positions
$\mathbf{r}_{1,2}$. The separation between these points, $\mathbf{m} =
\mathbf{r}_1 - \mathbf{r}_2$, will be called the dipole moment. Substituting
Eq.\ (\ref{current}) into Eq.\ (\ref{SL_phase}), we find the action for the
dipole:
\begin{equation}
 S_m
  = 2 S_v + 2 K \ln(\Lambda |\mathbf{m}|).
 \label{Sm}
\end{equation}
Here $\Lambda$ is the ultraviolet cut-off of the order of inverse radius of the
vortex core. The action $2S_v$ associated with two vortex cores is explicitly
added. We see that the action of the dipole grows logarithmically with
increasing $m$. This is true provided the dipole is smaller than the infrared
length $L$. For larger dipoles, the mass term will modify the dipole saddle
point, and the action will grow linearly with $m$. Thus we conclude that the 
infrared cut-off $L$ in the $\tilde Q$ theory effectively limits the maximum
allowed size of a dipole.

Let us now consider a general situation with arbitrary number of replicas. In
class AIII, we place the dipole in the first replica and represent the matrix
$\tilde Q$ as
\begin{equation}
 \tilde Q_m
  = \mathbbm{1} + |p\rangle (e^{i\phi} - 1) \langle p|.
 \label{Qm}
\end{equation}
Here $|p\rangle = \{1, 0, 0, \dots\}$ is the $N$-component vector in the replica
space. We can generate other dipole configurations with the same action
(\ref{Sm}) by moving and rotating the dipole in the real space, which changes
just the function $\phi$, and also by rotations $\tilde Q_m \mapsto \bar V
\tilde Q_m V$ in replica space with spatially constant unitary matrices $V$ and
$\bar V$. The mass term in Eq.\ (\ref{SL}) requires $\tilde Q_m = \mathbbm{1}$
at infinity and thus restricts the matrices by $\bar V V = 1$. Such unitary
rotations lead to the same form of $\tilde Q_m$ given by Eq.\ (\ref{Qm}) with an
arbitrary $N$-dimensional \emph{complex} unit vector $|p\rangle$. Thus
equivalent dipole configurations in replica space span the manifold
$\mathrm{CP}^{N-1} = \mathrm{U}(N) / \mathrm{U}(N-1) \times \mathrm{U}(1) =
S^{2N - 1} / S^1$.

In class CII, additional constraint $\bar V = V^T$ applies, see Eq.\
(\ref{bar}). As a result, allowed rotations generate an arbitrary unit
\emph{real} vector $|p\rangle$ with $N$ components. The corresponding space is
$\mathrm{RP}^{N-1} = \mathrm{O}(N) / \mathrm{O}(N - 1) \times \mathrm{O}(1) =
S^{N - 1}/S^0$. Finally, in class BDI, the constraint (\ref{bar}) leads to the
unit vector $|p \rangle$ composed of $N/2$ \emph{real quaternions}. The
corresponding manifold is $\mathrm{HP}^{N/2-1} = \mathrm{Sp}(N) /
\mathrm{Sp}(N - 2) \times \mathrm{Sp}(2) = S^{2N - 1}/S^3$. The spaces of
equivalent dipole configurations are listed in Table \ref{tab_D} together with
their volumes $V_p$. In all three chiral symmetry classes $V_p = O(1/N)$ in the
replica limit $N \to 0$.

\begin{table}
\tabcolsep=4pt
\begin{tabular}{cccc}
\hline\hline
Class & Null-space & $V_p$ \\
\hline
AIII & $\mathrm{CP}^{N-1}$ & $\pi^{N-1}/\Gamma(N)$ \\
CII  & $\mathrm{RP}^{N-1}$ & $\pi^{N/2-1}/\Gamma(N/2)$ \\
BDI  & $\mathrm{HP}^{N-1}$ & $\pi^{N-2}/\Gamma(N)$ \\
\hline\hline
\end{tabular}
\caption{List of spaces spanned by equivalent dipole configurations. Volumes of
these spaces are denoted by $V_p$.}
\label{tab_D}
\end{table}

\subsection{Derivation of RG equations}

The contribution of the dipole to the correlation functions $B_{1,2}$ of the
fast field $\tilde Q$  is found by substituting $\tilde Q = \tilde Q_m$ from
Eq.\ (\ref{Qm}) into Eqs.\ (\ref{B1}) and (\ref{B2}) and calculating the
Gaussian integral of quadratic fluctuations around the dipole saddle point. The
result has the form
\begin{equation}
 B_{1,2}^\text{dip}
  = \Lambda^4 \int d^2R\, d^2m\,
    \frac{Z'_\text{dip}}{Z_0}\; V_p\,
    \tilde B_{1,2}^\text{dip}\, e^{-S_m}.
 \label{Bdip}
\end{equation}
Here $S_m$ is the action (\ref{Sm}) associated with the dipole, $Z'_\text{dip}$
and $Z_0$ are the partition functions of quadratic fluctuations near the dipole
configuration and near the trivial uniform minimum of the action, respectively.
Zero modes related to rotations of the dipole in replica space are excluded from
the determinant $Z'_\text{dip}$; they yield the volume $V_p$ of the space of
equivalent dipoles, see Table \ref{tab_D}. Zero-modes related to the position
$\mathbf{R}$ of the center of mass of the dipole and nearly-zero modes related
to the dipole size $m$ are excluded from $Z'_\text{dip}$ as well and are taken
into account in Eq.\ (\ref{Bdip}) via integration over the corresponding
collective coordinates. The factor $\Lambda^4$ arises when the partition sum
over positions of vortex and antivortex, with the core size $\sim 1/\Lambda$, is
replaced by the integral over $\mathbf{R}$ and $\mathbf{m}$. (Appearance of this
factor is clear on dimensional reasons.) Finally, the pre-exponential factors
$\tilde B_{1,2}^\text{dip}$ are given by expressions inside angular brackets in
Eqs.\ (\ref{B1}) and (\ref{B2}) evaluated on $\tilde Q_m$ from Eq.\ (\ref{Qm}). 

The two vortices forming the dipole are located at $\mathbf{R} \pm
\mathbf{m}/2$. From Eqs.\ (\ref{current}) and (\ref{Qm}), we obtain
\begin{multline}
 \tilde Q_m^{-1} \nabla_\mu \tilde Q_m
  = -i \nabla_\mu \phi(\mathbf{r})\; | p \rangle \langle p | \\
  = -i \epsilon_{\mu\nu} \nabla_\nu \ln \left|
      \frac{\mathbf{r} - \mathbf{R} - \mathbf{m}/2}
           {\mathbf{r} - \mathbf{R} + \mathbf{m}/2}
    \right|\; | p \rangle \langle p |.
 \label{QnablaQ}
\end{multline}
In order to calculate $\tilde B_{1,2}^\text{dip}$, we substitute this expression
into Eqs.\ (\ref{B1}) and (\ref{B2}). Afterwards, Eq.\ (\ref{Bdip}) is used to
find $B_{1,2}^\text{dip}$. The projective property of the matrix (\ref{QnablaQ})
and the trace structure of Eqs.\ (\ref{B1}) and (\ref{B2}) imply the relation
\begin{equation}
 B_2^\text{dip}
  = u B_1^\text{dip},
 \label{u}
\end{equation}
where we have introduce the notation $u =  \mathop{\mathrm{Tr}} | p \rangle
\langle p|$. In the classes
AIII and CII, $u = 1$, while in the class BDI we have $u = 2$. The quantity $u$
can be interpreted as the minimal value of the number of replicas $N$ allowed by
the symmetry.

In view of the identity (\ref{u}), it suffices to calculate $B_1^\text{dip}$.
When the dipole position is fixed, the integrand in Eq.\ (\ref{B1}) depends
explicitly on both $\mathbf{r}$ and $\mathbf{r}'$ rather than on their
difference only. The uniformity is restored after integration over $\mathbf{R}$
in Eq.\ (\ref{Bdip}). Thus we will carry out the integration over the dipole
positions first.

Equations (\ref{B1}), (\ref{Bdip}), and (\ref{QnablaQ}) combine into
\begin{multline}
 B_1^\text{dip}
  = -\frac{u\, V_p}{8\pi s}\; \Lambda^4 \int d^2p\,
    \frac{Z'_\text{dip}(p)}{Z_0}\; e^{-S_m} \\
    \times \!\!\int\!\! d(\mathbf{r - r'}) \, d^2R\,
      \nabla \ln \left|
      \frac{\mathbf{r} - \mathbf{R} - \frac{\mathbf{m}}{2}}
           {\mathbf{r} - \mathbf{R} + \frac{\mathbf{m}}{2}}
      \right|
      \nabla' \ln \left|
      \frac{\mathbf{r}' - \mathbf{R} - \frac{\mathbf{m}}{2}}
           {\mathbf{r}' - \mathbf{R} + \frac{\mathbf{m}}{2}}
      \right|.
 \label{B1dip}
\end{multline}
By construction, gradients in Eq.\ (\ref{B1dip}) act on $\mathbf{r}$ and
$\mathbf{r}'$. Equivalently, we can assume that both gradients act on
$\mathbf{R}$. Integrating by parts and using the identity $\nabla^2
\ln|\mathbf{R}| = 2\pi \delta(\mathbf{R})$, we get rid of the $\mathbf{R}$
integral. Subsequent integration over $\mathbf{r} - \mathbf{r}'$ yields
\begin{equation}
 B_1^\text{dip}
  = -\frac{u \pi^2}{2 s}\; V_p \Lambda^4 \int\limits_{\Lambda^{-1}}^L dm\, m^3\,
    \frac{Z'_\text{dip}(p)}{Z_0}\; e^{-S_m}.
\end{equation}

Let us now analyze the determinant factor $Z'_\text{dip}/Z_0$. The major part of
the excitations near the dipole minimum of the action have wave lengths in the
interval between $\Lambda^{-1}$ and $m$. These modes, living in the relatively
slow ``background field'' created by vortices, lead to perturbative
renormalization of the parameters entering the dipole action $S_m$, in
accordance with Eqs. (\ref{RGsigma}) and (\ref{RGc}). The rest of the
determinant provides a non-universal factor dependent on the details of the
ultraviolet regularization or, equivalently, on the inner structure of the
vortex core. We do not assume any particular ultraviolet cutoff but instead will
include all non-universal factors in the definition of the vortex fugacity $y$. 
Using Eq.\ (\ref{Sm}), we represent the dipole contribution to $B_1$ in the
following form:
\begin{equation}
 B_1^\text{dip}
  \propto -V_p e^{-2 S_v} \Lambda
    \int\limits_{\Lambda^{-1}}^L dm\; (\Lambda m)^{3 - 2 K(m)}.
\end{equation}
All numerical factors omitted in this expression remain finite in the replica
limit $N \to 0$.

In order to find the dipole contribution to RG equations for $\sigma$ and $c$,
we express the derivative of $B_1^\text{dip}$ in terms of fugacity,
\begin{equation}
 \frac{\partial B_1^\text{dip}}{\partial \ln L}
  = -\frac{N y^2}{u\, \sigma}.
\end{equation}
All uncontrolled numerical prefactors are now hidden in the definition of $y$:
\begin{equation}
 y(L)
  \propto \sqrt{V_p/N}\, e^{-S_v}\, (\Lambda L)^{2 - K(L)}.
 \label{y}
\end{equation}
Note that the ratio $V_p/N$ remains finite in the replica limit. Taking
derivatives of the Eqs.\ (\ref{sigma}) and (\ref{c}) and using the identity
(\ref{u}), we readily obtain the dipole contribution to the RG equations in
terms of fugacity $y$. Adding the perturbative corrections from Eqs.\
(\ref{RGsigma}) and (\ref{RGc}), we get the result
\begin{align}
 \frac{\partial \sigma}{\partial \ln L}
  &=-N - \begin{cases}
      \dfrac{\sigma y^2}{1 + N}, & \text{AIII, BDI}, \\[6pt]
      \dfrac{\sigma y^2}{1 + N/2}, & \text{CII},
    \end{cases}
 \label{RGsigmaN} \\
 \frac{\partial c}{\partial \ln L}
  &=1 - cy^2 \left( 2 + \frac{N c}{\sigma} \right) - \begin{cases}
      \dfrac{\sigma y^2}{1 + N}, & \text{AIII, BDI}, \\[6pt]
      \dfrac{\sigma y^2}{2 + N}, & \text{CII}.
    \end{cases}
 \label{RGcN}
\end{align}
Renormalization of fugacity follows from Eq.\ (\ref{y}). Differentiating with
respect to $\ln L$ and then taking the limit $\Lambda L \to 1$, we obtain
\begin{equation}
 \frac{\partial y}{\partial \ln L}
  = (2 - K) y.
 \label{RGy}
\end{equation}
This equation is expressed in terms of the stiffness parameter $K$ defined in
Eq.\ (\ref{K}). Renormalization of $K$ follows from the other two RG equations
(\ref{RGsigmaN}) and (\ref{RGcN}). In the replica limit $N \to 0$, the
equations for $\sigma$ and $K$ acquire the form
\begin{align}
 \frac{\partial \sigma}{\partial \ln L}
  &=-\sigma y^2,
 \label{RGsigma_final} \\[6pt]
 \frac{\partial K}{\partial \ln L}
  &=\begin{cases}
      \dfrac{1}{4} - 2 K y^2, & \text{AIII}, \\[6pt]
      \dfrac{1}{8} - 2 K y^2 + \dfrac{\sigma y^2}{16}, & \text{CII}, \\[6pt]
      \dfrac{1}{2} - 2 K y^2 - \dfrac{\sigma y^2}{4}, & \text{BDI}.
    \end{cases}
 \label{RGK}
\end{align}

From this result we see that vortices provide negative localizing correction to
the conductivity. In a good metallic sample with $\sigma \gg 1$ and
exponentially small (in $\sigma$) fugacity $y$, vortices are bound in tiny
dipoles. The overall negative correction to conductivity remains finite at long
length scales since the fugacity rapidly decreases in the course of
renormalization. With lowering the starting value of conductivity and increasing
fugacity, larger dipoles come into play and negative correction to $\sigma$ is
more pronounced. When both $\sigma$ and $y$ are of order 1, the phase transition
occurs. Our RG equations are not quantitatively accurate in this limit but can
be used for a qualitative description of the transition.

\begin{figure}
\begin{center}
\includegraphics[width=0.9\columnwidth]{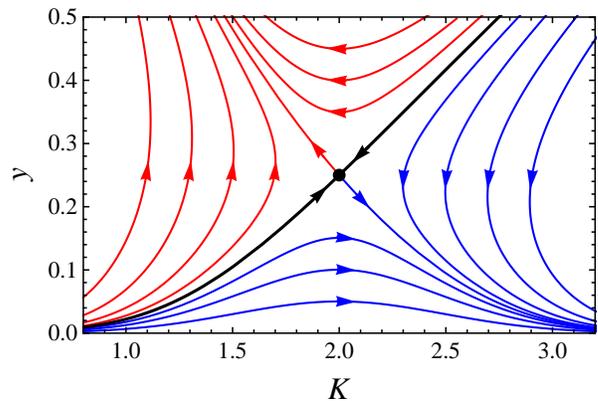}
\end{center}
\caption{Schematic RG flow near the metal-insulator transition in the symmetry
class AIII according to Eqs.\ (\ref{RGy}) and (\ref{RGK}).}
\label{flow}
\end{figure}

\subsection{Vortex-driven localization}

Let us study in more detail the RG equations in class AIII. As was mentioned in
the end of Sec.\ \ref{sec:perturb}, Eqs.\ (\ref{RGsigma_final}) and (\ref{RGK})
are exact to all orders in the $1/\sigma$ loop expansion. The only assumption
used in deriving these equations is $y \ll 1$. Let us neglect this assumption
for a moment and treat the RG equations as if they were exact. Another feature
of the symmetry class AIII is the partial separation of variables.
Renormalization of $K$ and $y$ is independent of $\sigma$. This allows us to
study the 2D flow in the $K$ -- $y$ plane. The flow is depicted in Fig.\
\ref{flow}. The diagram contains one unstable fixed point at the position $K =
2$, $y = 1/4$. The black line passing through this point separates the two
stable phases of the system. Below this line, the flow is directed towards $y =
0$ and $K \to \infty$. This part of the diagram corresponds to metallic phase.
With vanishing fugacity, renormalization of conductivity stops at some finite
value, while the stiffness $K$ keeps increasing solely due to renormalization of
the Gade parameter $c$. In the metallic phase, the system rapidly approaches the
regime described by the Gade sigma model without vortices. The finite
contribution of the dipoles can be included into bare parameters of this sigma
model.

Above the critical line, fugacity $y$ increases indefinitely, which signifies
vortex unbinding. This phase is analogous to the high-temperature disordered
phase in the BKT problem and corresponds to the insulating state of the sample.
Indeed, according to Eq.\ (\ref{RGsigma_final}), the conductivity flows all the
way to zero with growing $y$. Perturbative RG equations rapidly become
inaccurate in this limit and cannot be used deep in the insulating phase.
Nevertheless, using analogy with the BKT transition, we conclude that the
distinction between metal and insulator states is robust.

RG equations (\ref{RGy}), (\ref{RGsigma_final}), and (\ref{RGK}) were derived
assuming small value of the fugacity (and large value of $\sigma$ in classes
CII and BDI). This means that the metal-insulator transition fixed point lies
at the border of the applicability region of these RG equations. The theory is
not quantitatively controllable in the vicinity of the fixed point but the
overall qualitative picture of the RG flow and, in particular, the very presence
of the unstable fixed point should be correct. Numerical simulations of electron
transport in 2D chiral disordered systems would provide a quantitative estimate
of the critical properties of the localization transition.

We have found a fixed point of the metal-insulator transition at $\sigma = 0$.
This signifies the absence of minimal metallic conductivity in the considered
problem. In a realistic disordered chiral system, the bare value of fugacity is
related to the Drude conductivity by an exponential law $\ln y \propto -\sigma$.
This means that the critical state is achieved when both $y$ and $\sigma$ are of
order unity. Hence the Anderson transition will normally happen at $\sigma \sim
1$. The exact critical value of $\sigma$ depends on microscopic details of a
particular realization of the disordered system.

In the next section we present an alternative framework for the derivation of
non-perturbative renormalization equations for the chiral class AIII. (We expect
that a generalization of this approach to other chiral classes is possible as
well but it requires further work.)

\section{Class AIII and generalized sine-Gordon theory}
\label{sec:sine-gordon}

\subsection{Sigma-model and duality transformation}
\label{subsec:duality}

The group nature of the sigma-model target space for the class AIII allows us to
develop yet another description of vortices, bearing analogy with the
sine-Gordon theory of the BKT transition.\cite{Minnhagen} To construct such a
description we first work out a dual representation of the AIII sigma model. The
inclusion of vortex excitations turns out to be straightforward within this
approach.   

As a starting remark, let us note that for any field $Q(\mathbf{r})$ of unitary
matrices one can define an associated two-component vector field  
\begin{equation}
 \mathbf{h}
  = -i Q^{-1} \nabla Q.
\end{equation}
The components of $\mathbf{h}$  are Hermitian $N \times N$ matrices. Conversely,
given a vector field $\mathbf{h}$ with Hermitian entries and the value of the
unitary matrix $Q$ at a single point, one can reconstruct $Q$ in the whole plane
by integrating the system of equations 
\begin{equation}
 \nabla_\mu Q
  = i Q h_\mu.
\label{Q_h_correspondence}
\end{equation}
The solution of Eq.\ (\ref{Q_h_correspondence}), if exists, is automatically a
unitary matrix field. Examining the cross-derivatives of $Q$, we find that the
compatibility condition for the system (\ref{Q_h_correspondence}) is given by 
\begin{equation}
 F_{\mu\nu}
  \equiv \nabla_\mu h_\nu - \nabla_\nu h_\mu + i [h_\mu, h_\nu]
  = 0. 
 \label{integrability_1}
\end{equation}
Quite remarkably, the integrability constraint (\ref{integrability_1}) can be
viewed as a requirement of vanishing field-strength tensor for the non-abelian 
``gauge field'' $\mathbf{h}$. 

We thus come to the conclusion that the matrix field $Q$ is in one-to-one
correspondence with the vector field $\mathbf{h}$ subject to the integrability
constraint (\ref{Q_h_correspondence}). In particular, the grand partition
function of the sigma model can be represented as
\begin{equation}
 Z
  = \int DQ\, e^{-S[Q]}
  = \int \left\{
      \prod_\mathbf{r} d^2h(\mathbf{r})\, \delta\big[ F_{xy}(\mathbf{r}) \big]
    \right\} e^{-S[\mathbf{h}]},
 \label{int_dQ_dh}
\end{equation}
where the action functional is quadratic in $\mathbf{h}$, cf.\ Eq.\
(\ref{action}):
\begin{equation}
 S[\mathbf{h}]
  = \int \frac{d^2 r}{8\pi} \left[
      \sigma \mathop{\mathrm{Tr}} \mathbf{h}^2
      +c\, (\mathop{\mathrm{Tr}} \mathbf{h})^2
    \right]. 
\end{equation}
A straightforward inspection shows that the integration measure over
$\mathbf{h}$ in Eq.\ (\ref{int_dQ_dh}) is indeed flat.

The functional $\delta$-function in Eq.\ (\ref{int_dQ_dh}) can be resolved by
the integration over an auxiliary Hermitian matrix field $\Theta$ 
\begin{align}
 Z
  &= \int D \mathbf{h}\, D\Theta\, e^{-S[\mathbf{h}, \Theta]},
 \label{int_dQ_dh_dtheta} \\
 S[\mathbf{h}, \Theta]
  &= S[\mathbf{h}] + i \int d^2r \mathop{\mathrm{Tr}} (\Theta F_{xy}).
 \label{int_dQ_dh_dtheta_action}
\end{align}
Since the action $S[\mathbf{h}, \Theta]$ is quadratic in $\mathbf{h}$, we can
easily perform the integration over $\mathbf{h}$. As a result, we obtain a dual
representation of the AIII sigma model in terms of the matrix $\Theta$:
\begin{align}
 Z
  &= \int D\mu(\Theta) e^{-S[\Theta]}, \label{Z_theta} \\
 S[\Theta]
  &= \frac{2\pi}{\sigma + N c} \int d^2r\, (\nabla \theta_0)^2 \notag \\
  &+ \frac{2\pi}{\sigma} (\delta_{\mu\nu} + i\epsilon_{\mu\nu})
     \int d^2r\, (\nabla_\mu \theta_a) g_{ab}(\theta) (\nabla_\nu \theta_b). 
  \label{S_theta}
\end{align}
Here we have introduced a decomposition $\Theta = \theta_0/\sqrt{N} +
\sum_{a=1}^{N^2-1} \theta_a T^a$ of the matrix $\Theta$ over the generators of
$\mathrm{U}(N)$. Further, $g_{ab}$ is a $(N^2 - 1) \times (N^2 - 1)$ matrix
given by 
\begin{equation}
 g
  = \big( \mathbbm{1} - 4\pi Z/\sigma \big)^{-1},
 \qquad
 Z_{ab}
  = \sum_{c = 1}^{N^2 - 1} \theta_c f_{abc},
\end{equation}
where $f_{abc} = -i\mathop{\mathrm{Tr}} T^a [T^b, T^c]$ are the structure
constants of the $\mathfrak{su}(N)$ algebra. The measure of functional
integration over $\Theta$ in (\ref{Z_theta}) is $D\mu(\Theta) =
\prod_\mathbf{r} \det\{g[\Theta(\mathbf{r})]\}\, d\Theta(\mathbf{r})$. Since $(g
+ g^T)/2 = g g^T$ can be regarded as the metric tensor of the model, we
recognize that the integration measure is consistent with the metric. 

By construction, the model defined in Eqs.\ (\ref{Z_theta}), (\ref{S_theta})
is fully equivalent to the original AIII sigma model. In particular, the two
theories should obey identically the same perturbative renormalization group. In
Appendix \ref{app:theta_renormalization} we verify this fact explicitly within a
one-loop calculation.  

So far we were completely neglecting the vortex excitations. To include them
into the theory, one must realize that integrability condition Eq.\
(\ref{integrability_1}) can be slightly relaxed. Indeed, as discussed above, a
single vortex centered at the origin of the coordinate frame is described by the
$Q$ matrix 
\begin{equation}
 Q
  = \mathbbm{1} + |p\rangle (e^{i\phi} - 1) \langle p|.
\end{equation}
Here $\phi$ is the polar angle and $|p\rangle \langle p|$ projects onto some
unit vector $|p\rangle$ in the replica space. The corresponding $\mathbf{h}$
field has only the azimuthal component $h_\phi = |p\rangle \langle p|/r$ and the
field-strength tensor assumes the form
\begin{equation}
 F_{xy}(\mathbf{r})
  = 2\pi\delta(\mathbf{r}) |p\rangle \langle p|.
 \label{F_single}
\end{equation}
A straightforward generalization of Eq.\ (\ref{F_single}) to the case of an
arbitrary collection of vortices and anti-vortices located at points
$\mathbf{r}_i$ and characterized by projectors $|p_i\rangle \langle p_i|$ reads 
\begin{equation}
 F_{xy}(\mathbf{r})
  = 2\pi\sum_i \eta_i\delta(\mathbf{r} - \mathbf{r}_i) |p_i \rangle\langle p_i|,
 \qquad
 \eta_i
  = \pm 1. 
 \label{F_vortices}
\end{equation}

We now repeat the treatment leading to the construction of the dual sigma-model
representation (\ref{Z_theta}), (\ref{S_theta}) but this time taking into
account vortex configurations [i.e. replacing the constraint $F_{\mu\nu} = 0$ by
Eq.\ (\ref{F_vortices}). This yields
\begin{multline}
 Z
  = \int D\mu(\Theta) e^{-S[\Theta]}
    \sum_{n = 1}^\infty \frac{(\Lambda^2 y_0)^n}{n!} \\
    \times \sum_{\eta_i=\pm1} \prod_{i = 1}^n \int d^2 r_i\, d|p_i\rangle
    \exp[2\pi i\eta_i \langle p_i| \theta(\mathbf{r}_i) |p_i \rangle] \\
  =  \int D\mu(\Theta) e^{-S[\Theta] - S_y[\Theta]}.
\end{multline}
Here $y = e^{-S_v}$ is the bare value of fugacity, integration with respect to
$|p_i \rangle$ is performed over all complex vectors of unit length with natural
integration measure, and the vortex-induced correction to the action reads
\begin{equation}
 S_y[\Theta]
  \equiv -2 y \Lambda^2 \int d^2r\, d|p \rangle\,
    \cos 2\pi \langle p|\Theta(\mathbf{r})|p \rangle. 
 \label{Sy}
\end{equation}
The action of the dual theory given by a sum of Eqs.\ (\ref{S_theta}) and
(\ref{Sy}) constitutes the central result of this section. It provides a
convenient starting point for the generation of the RG equations (including
non-perturbative contributions) for the sigma model of class AIII.

\subsection{RG analysis}
\label{subsec:rg}

We are now in a position to explore the RG flow of the AIII sigma model from the
point of view of the dual representation. In the minimal model of class AIII,
the $\mathrm{U}(1)$ sigma model, Eqs.\ (\ref{S_theta}) and (\ref{Sy}) reduce to
the standard sine-Gordon action
\begin{equation}
 S[\Theta] + S_y[\Theta]
  = \int d^2r \left[
      \frac{2\pi}{\sigma + c}(\nabla \theta_0)^2
      -2 y \Lambda^2 \cos 2\pi \theta_0
    \right], 
\end{equation}
and we recover the BKT renormalization group. Let us now explore the
contribution of vortices to RG equations in a general situation $N \neq 1$. As
in the BKT theory, we will do this perturbatively in the vortex fugacity $y_0$.
We neglect for a while the non-linear terms in  $S[\Theta]$, i.e. replace
$S[\theta]$ by 
\begin{equation}
 S_0[\theta]
  = 2\pi \int d^2r \left[
      \frac{(\nabla \theta_0)^2}{\sigma + N c}
      + \sum_a \frac{(\nabla \theta_a)^2}{\sigma}
    \right].
 \label{S_0}
\end{equation}
Following the standard procedure we decompose the fields $\Theta$ into
the fast and slow modes $\Theta = \Theta_< + \Theta_>$, with the fast modes
$\Theta_>$ populating a thin shell in momentum space $\Lambda' \equiv \Lambda -
\Delta\Lambda < |k| < \Lambda$. The action functional becomes
\begin{equation}
 S[\Theta_> + \Theta_<]
  = S_0[\Theta_<] + S_0[\Theta_>] + S_y[\Theta_<] + S_1[\Theta_<, \Theta_>],
\end{equation}  
where
\begin{align}
 S_1[\Theta_>, \Theta_<]
  &= S_c[\Theta_>, \Theta_<] + S_s[\Theta_>, \Theta_<], \\
 S_{c(s)}[\Theta_>, \Theta_<]
  &= -2\Lambda^2 y \int d^2r\, d|p\rangle\, U_{c(s)}(\mathbf{r}, |p\rangle), \\
 U_c(\mathbf{r}, |p\rangle)
  &= \big(\!\cos 2\pi \langle p|\Theta_>(\mathbf{r}) |p\rangle - 1\big) \notag\\
  &\qquad\times \cos 2\pi \langle p|\Theta_<(\mathbf{r}) |p\rangle, \\
 U_s(\mathbf{r}, |p\rangle)
  &= -\sin 2\pi \langle p|\Theta_>(\mathbf{r}) |p\rangle
     \sin 2\pi \langle p|\Theta_<(\mathbf{r}) |p\rangle.
\end{align}

It is easy to see that to the first order in $y$ the interaction term $S_1$
generates the correction to fugacity itself governed by the RG equation [the
stiffness parameter $K$ was defined in Eq.\ (\ref{K})]: 
\begin{equation}
 \frac{\partial y}{\partial \ln \Lambda/\Lambda'}
  = (2 - K) y.
 \label{sine_gordon_fugacity_rg}
\end{equation}

We now turn to terms of the second order in fugacity. We observe that, strictly
speaking, the action of our model does not preserve its form under the RG
transformations. For example, let us consider the contribution to the action due
to the interaction $S_{c}$: 
\begin{multline}
 \Delta S_c[\Theta_<]
  = -\frac{1}{2} \big<\!\big< S_c^2 \big>\!\big>
  = -2 \Lambda^4 y^2 \int d^2R\, d^2r\, d|p_1\rangle\, d|p_2\rangle\, \\
  \times \big<\!\big<
      U_c(\mathbf{R}_+, |p_1\rangle) U_c(\mathbf{R}_-, |p_2\rangle)
    \big>\!\big>, 
\end{multline}
where $\mathbf{R}_{\pm} = \mathbf{R} \pm \mathbf{r}/2$. The gradient expansion
of the slow fields leads to
\begin{multline}
 \Delta S_c[\Theta_<]
  = -\Lambda^4 y^2 \int d^2R\, d^2r\, d|p_1\rangle d|p_2\rangle\,
    (X_+ + X_-), \\
  \times \big<\!\big<
      \cos 2\pi \langle p_1|\Theta_>(\mathbf{R}_+) |p_1\rangle
      \cos 2\pi \langle p_2|\Theta_>(\mathbf{R}_-) |p_2\rangle
    \big>\!\big>,
\label{DeltaS_c_theta<}
\end{multline}
with
\begin{multline}
 X_\pm
  = \cos 2\pi \big[
      \langle p_1|\Theta_<(\mathbf{R}) |p_1\rangle
      \pm \langle p_2|\Theta_<(\mathbf{R}) |p_2\rangle
    \big]\\ 
  \times \left[
      1 \pm \frac{r^2}{4} \langle p_1|\nabla \Theta_<(\mathbf{R})|p_1\rangle 
      \langle p_2|\nabla\Theta_<(\mathbf{R}) |p_2\rangle
    \right].
\end{multline}
We see that the RG flow generates terms of the form $\cos 2\pi\big[ \langle
p_1|\Theta_< |p_1\rangle \pm \langle p_2|\Theta_< |p_2\rangle \big]$
corresponding to the creation of two vortices (for plus sign) or a vortex and
anti-vortex (for minus sign) sitting at the same point and characterized by the
projectors $|p_1\rangle \langle p_1|$ and $|p_2\rangle \langle p_2|$. While the
former process is irrelevant in RG sense, the latter one can be even more
relevant perturbation than the initial $\cos 2\pi \langle p|\Theta_< |p\rangle$
term if vectors $|p_1\rangle$ and $|p_2\rangle$ are sufficiently close. However,
these terms are suppressed by additional power of fugacity $y\ll 1$ as compared
to the original vortex term, and we neglect them. Clearly, this neglect is
not justified in the region $y \sim 1$ where the fixed point governing the
transition is located. At the same time, the theory we are developing is
quantitatively controllable only at $y \ll 1$. Discarding the above terms should
not lead to any qualitative changes in the RG flow. 

We are only interested in the most important $y^2$ terms, i.e., those that
produce contributions to renormalization of $\sigma$ and $K$ and are thus
responsible for the localization transition. To obtain them, we approximate the
correction (\ref{DeltaS_c_theta<}) to the action of the slow fields by
\begin{multline}
 \Delta S_c[\Theta_<]
  = 2\pi \kappa \Lambda^4 y^2\int d^2R\, d|p\rangle\,
    \langle p|\nabla\Theta_<(R) |p\rangle^2 \\ 
  \times \!\!\int\!\! d^2r\, r^2\, \big<\!\big<\!
      \cos 2\pi \langle p|\Theta_>(\mathbf{R}_+) |p\rangle
      \cos 2\pi \langle p|\Theta_>(\mathbf{R}_-) |p\rangle
    \big>\!\big>.
\end{multline}
Here $\kappa$ is a numerical coefficient. Averaging now over the fast fields and
singling out the contributions of the first order in $\Delta\Lambda$ we get 
\begin{equation}
 \Delta S_c[\Theta_<]
  = 2\pi \kappa \frac{\Delta\Lambda}{\Lambda} K^2 y^2 \int d^2R\, d|p\rangle\,  
    \langle p|\nabla\Theta_<(\mathbf{R})|p\rangle^2. 
\end{equation}
(We have absorbed an additional numerical coefficient into  $\kappa$.) Noting
that for arbitrary matrices $A$ and $B$
\begin{equation}
 \int d|p\rangle \langle p|A|p\rangle \langle p|B|p\rangle
  = \frac{V_p}{N(N + 1)} \big(
      \mathop{\mathrm{Tr}} A \mathop{\mathrm{Tr}} B + \mathop{\mathrm{Tr}} A B
    \big), 
\end{equation}
we finally get
\begin{multline}
 \Delta S_c[\Theta_<]
  = 2\pi \kappa \frac{\Delta\Lambda}{\Lambda} \frac{V_p}{N(N+1)} K^2 y^2\\
  \times \left[
      (n+1) (\nabla\theta^0_<)^2 + \sum_{a = 1}^{N^2 - 1} (\nabla\theta^a_<)^2
    \right]. 
 \label{delta_Sc_final}
\end{multline}
Comparing now Eq.\ (\ref{delta_Sc_final}) with the Gaussian action (\ref{S_0}),
we conclude that integration over the fast modes has generated corrections to
$\sigma$ and $c$ given by 
\begin{align}
 \Delta\sigma
  &= -\kappa \frac{y^2 K^2 \sigma^2}{\Gamma(N+2)}
     \frac{\Delta\Lambda}{\Lambda}, \label{corrections1} \\ 
 \Delta c
  &= -\kappa \frac{y^2 K^2}{\Gamma(N+1)} \frac{\Delta\Lambda}{\Lambda} \left[
      \frac{\sigma^2}{N + 1} + 2 \sigma c + N c^2
    \right]. \label{corrections2}
\end{align}

So far we were neglecting non-Gaussian terms in the action $S[\Theta]$. When
taken into account, they will induce the perturbative renormalization of the
parameters of the model and, in particular, lead to the appearance of the
scale-dependent $\sigma$ and $c$ in Eqs.\ (\ref{corrections1}) and
(\ref{corrections2}). Apart from this, the higher order terms do not affect our
analysis: a straightforward calculation shows that in our approximation the
interaction term $\Delta S_s$ does not produce corrections of order $y^2$.
Combining Eqs.\ (\ref{sine_gordon_fugacity_rg}), (\ref{corrections1}), and
(\ref{corrections2}), we obtain a system of RG equations for the AIII sigma
model
\begin{align}
 \frac{\partial \sigma}{\partial \ln L}
  &= -N - \frac{y^2 K^2 \sigma^2}{\Gamma(N+2)},\\ 
 \frac{\partial K }{\partial \ln L}
  &= \frac{1 - N}{4} - \frac{y^2 K^2}{\Gamma(N+1)} \left(
      \frac{2\sigma^2}{N + 1} + 2 \sigma c + N c^2
    \right),\\ 
 \frac{\partial y}{\partial \ln L}
  &= (2 - K) y.
\end{align}
These equations are equivalent to those obtained in Sec.\ \ref{sec:nonperturb}
up to subleading terms. To make this equivalence apparent, we redefine the
fugacity parameter according to $y \mapsto (y / K) \sqrt{\Gamma(N + 1)/\sigma}$.
This amounts to changing the (uncontrolled) pre-exponential factor in the
exponentially small quantity $y$. After such a rescaling, the RG equations
(\ref{RGsigmaN}), (\ref{RGcN}), and (\ref{RGy}) are reproduced.

\section{Topological terms}
\label{sec:topology}

In previous sections, we discussed localization within the Gade-Wegner sigma
model, Eq.\ (\ref{action}), of three chiral classes. Now we will consider the
situations when the sigma-model action is augmented with an additional
topological term. The topology of the sigma-model target spaces allows (in two
dimensions) inclusion of the Wess-Zumino term in class AIII and the
$\mathbb{Z}_2$ $\theta$ term in class CII. We will demonstrate below that these 
extra terms (arising in models of random Dirac fermions of the corresponding
symmetries) crucially affect localization properties. Specifically, we show
that vortex excitations do not appear when a topological term is present.

\subsection{Class AIII with Wess-Zumino term}
\label{sec:topologyaiii}

The field of the sigma model of class AIII is a unitary matrix $Q(\mathbf{r})
\in \mathrm{U}(N)$, with $\mathbf{r}$ belonging to the 2D coordinate space. In
our analysis we consider only field configurations with the $Q$ matrix taking
some fixed value at spatial infinity. (Otherwise the sigma-model action
inevitably diverges due to gradient terms.) This allows us to compactify the
coordinate space making it equivalent to a two-sphere. The compactified real
space $S^2$ can be viewed as a surface of a three-dimensional solid ball and we
introduce the radial coordinate $\tau$ such that $\tau = 1$ at the surface
(i.e., in the physical 2D space) and $\tau = 0$ in the center of the ball. The
matrix $Q$ can be continuously extended to the interior of the ball such that
\begin{equation}
 Q(\mathbf{r}, \tau)
  = \begin{cases}
      Q(\mathbf{r}), & \tau = 1,\\
      \mathop{\mathrm{const}}, & \tau = 0.
    \end{cases}
 \label{Qext}
\end{equation}
This extension is always possible since the second homotopy group of the target
space is trivial, $\pi_2(U) = 0$.

In terms of the extended matrix $Q(\tau)$, the Wess-Zumino term acquires the
form
\begin{multline}
 S_\text{WZ}[Q]
  = \frac{i k}{12\pi}\, \epsilon_{\mu\nu\lambda} \\
    \times \int d\tau\, d^2r \mathop{\mathrm{Tr}} \Big(
      Q^{-1} \nabla_\mu Q Q^{-1} \nabla_\nu Q Q^{-1} \nabla_\lambda Q
    \Big).
 \label{WZ}
\end{multline}
The integrand in this expression explicitly depends on the values of $Q$ at
$\tau \neq 1$, i.e., away from the physical 2D space. However, the variation of
the Wess-Zumino action can be represented as an integral of a three-dimensional
vector divergence:
\begin{multline}
 \delta S_\text{WZ}[Q]
  = \frac{i k}{4\pi}\, \epsilon_{\mu\nu\lambda} \\
    \times \int d\tau\, d^2r \nabla_\mu \mathop{\mathrm{Tr}} \Big(
      Q^{-1} \delta Q Q^{-1} \nabla_\nu Q Q^{-1} \nabla_\lambda Q
    \Big) \\
  = \frac{i k}{4\pi}
    \int d^2r \mathop{\mathrm{Tr}} \Big(
      Q^{-1} \delta Q \big[ Q^{-1} \nabla_x Q, Q^{-1} \nabla_y Q \big]
    \Big).
 \label{varWZ}
\end{multline}
Thus the actual value of the Wess-Zumino term is determined only by the physical
values of $Q$ at $\tau = 1$ up to a constant. This constant does not change with
small variations of $Q$ away from the physical 2D space but takes different
values for topologically distinct extensions of $Q$ in the third dimension.
These nonequivalent extensions are classified by the third homotopy group
$\pi_3(U) = \mathbb{Z}$. For any two extensions the values of the Wess-Zumino
term differ by $2\pi i k$ times an integer number. Thus the Wess-Zumino theory
is well-defined for any integer value of $k$.

Introducing vortices in the Wess-Zumino theory is problematic. In order to
avoid the singularity in the center of a vortex, we exclude a small region of
the vortex core from our physical space. This introduces a boundary in the
problem and $Q$ is not a constant along this boundary. As a result, the 2D
physical space cannot be compactified to the two-sphere. Thus the construction
of the Wess-Zumino term, involving extension to the third dimension, becomes
ill-defined.

One naive way to overcome this difficulty is to use a local 2D representation of
the Wess-Zumino term.\cite{Witten} As was discussed above, the Wess-Zumino term
actually depends only on the values of $Q$ in the physical space. Using any
explicit parameterization of the unitary matrix $Q$ by a set of coordinates
$\psi_i$, the Wess-Zumino term can be written as a 2D integral of a suitable
skew-symmetric differential form $\lambda_{ij}$:
\begin{equation}
 S_\text{WZ}[Q]
  = i k\, \epsilon_{\mu\nu} \int d^2r\; \lambda_{ij}
    \big[ \psi(\mathbf{r}) \big]  \nabla_\mu \psi_i \nabla_\nu \psi_j.
 \label{WZlocal}
\end{equation}
In this local representation, one can integrate the differential form over a
finite physical space with boundary. Such a construction is however
unsatisfactory because it violates the global gauge symmetry of the system.
Indeed, the Wess-Zumino action (\ref{WZ}) is manifestly invariant under the
global transformation $Q \mapsto U_L Q U_R$ parameterized by two constant
unitary matrices $U_{L,R}$. However, the local density of the Wess-Zumino term,
Eq.\ (\ref{WZlocal}), is not invariant under global rotation of fields. Instead,
the tensor $\lambda_{ij}$ transforms as
\begin{equation}
 \lambda_{ij}
  \mapsto \lambda_{ij} + \frac{\partial\beta_j}{\partial\psi_i}
    - \frac{\partial\beta_i}{\partial\psi_j}
\end{equation}
where a set of functions $\beta(\psi)$ encodes the information about $U_L$ and
$U_R$. The local expression for the Wess-Zumino term changes by an integral of a
total derivative:
\begin{equation}
 S_\text{WZ}
  \mapsto S_\text{WZ} + 2 i k\, \epsilon_{\mu\nu} \int d^2r\;
    \nabla_\mu(\beta_j \nabla_\nu \psi_j).
 \label{WZgauge}
\end{equation}
If the model is considered on a manifold without a boundary, the above integral
vanishes and the Wess-Zumino term is indeed invariant under the gauge
transformation. If, however, the real space integration is performed over a
bounded region, the integral in Eq.\ (\ref{WZgauge}) yields the circulation of
$\beta_j \nabla \psi_j$ along the boundary and may become non-zero. This
signifies the breakdown of the gauge symmetry at the boundary.

In order to understand better the boundary effects in the Wess-Zumino theory, we
will resort to the disordered fermion problem yielding the class AIII sigma
model with the Wess-Zumino term. The typical example is given by disordered
massless Dirac fermions with random vector potential. The Hamiltonian of such a
model has the form
\begin{equation}
 H
  = \begin{pmatrix}
      0 & p_x - i p_y + a \\
      p_x + i p_y + a^\dagger
    \end{pmatrix}.
 \label{HAIII}
\end{equation}
Here $a$ is a random complex-valued matrix acting in the auxiliary flavor space.
The only symmetry of the Hamiltonian is the chiral symmetry $H = -\sigma_z H
\sigma_z$. Since the spectrum of the Hamiltonian is unbounded, it appears to be
impossible to introduce the boundary condition (a classically forbidden region
for massless Dirac fermions) preserving the chiral symmetry. In order to model a
hole in the sample, one has to add an additional mass term $m \sigma_z$ to the
Hamiltonian and consider the limit of large $m$. Such a term explicitly breaks
the chiral symmetry. In the corresponding sigma model, the $Q$ matrix at the
hole boundary will be restricted to the manifold $\mathcal{M}_A = \mathrm{U}(N)
/ \mathrm{U}(N/2) \times \mathrm{U}(N/2)$ of class A. Such a boundary condition
will maintain only the diagonal part of the global gauge symmetry $U_L =
U_R^{-1}$. This also forbids the vortex excitations inside the hole since
$\mathcal{M}_A$ is simply connected. Thus we see that the short distance
regularization (making a small hole), needed to introduce a vortex in the sigma
model, breaks the chiral symmetry and does not allow a vortex excitation. We
conclude that vortices are incompatible with the Wess-Zumino term in the action
of class AIII.

The model of massless Dirac fermions in a random magnetic field, described by
the Hamiltonian (\ref{HAIII}) is exactly solvable. The coupling constant
characterizing the vector potential strength is exactly marginal, so that the
model possess a line of fixed points. A remarkable property of this problem is
that the system never gets truly localized, however strong the disorder is. In
particular, the conductivity is equal to $e^2/\pi h$ for any disorder.
Therefore, the absence of the localizing vortex contribution in the
corresponding
sigma-model is consistent with the known exact solution of the underlying
fermionic problem.

Let us illustrate the incompatibility of vortices and Wess-Zumino term in a
more explicit way. We will construct an extension of the sigma-model manifold
such that the theory will be well-defined inside the vortex core while all the
symmetries are preserved. Let us extend the $Q$ matrix by one row and one column
embedding $\mathrm{U}(N)$ into $\mathrm{SU}(N + 1)$. We will associate a mass
$M$ with the extra degrees of freedom suppressing them away from a vortex core.
Explicitly, consider the following action for the extended unitary matrix
$\underline{Q}$:
\begin{multline}
 S[\underline{Q}]
  = -\int \frac{d^2r}{8\pi} \Bigg[
      \sigma \mathop{\mathrm{Tr}}(\underline{Q}^{-1} \nabla \underline{Q})^2
      +M^2 \mathop{\mathrm{Tr}}(\underline{Q} R \underline{Q}^{-1} R) \\
      +\frac{c - \sigma}{4}
        (\mathop{\mathrm{Tr}} R \underline{Q}^{-1} \nabla \underline{Q})^2
    \Bigg] + S_\text{WZ}[\underline{Q}].
 \label{Sext}
\end{multline}
Here the matrix $R$ of the form
\begin{equation}
 R
  = \begin{pmatrix}
      -1 & 0 \\
      0 & \mathbbm{1}_{N \times N}
    \end{pmatrix}
\end{equation}
is introduced to single out the off-diagonal elements in the first row and
first column of $\underline{Q}$. These elements are made massive by the second
term of the action (\ref{Sext}).

Away from vortices the matrix $\underline{Q}$ contains only the soft modes
arranged as follows:
\begin{equation}
 \underline{Q}_\text{soft}
  = \begin{pmatrix}
      \det Q^{-1} & 0 \\
      0 & Q
    \end{pmatrix}.
 \label{uQsoft}
\end{equation}
The lower right diagonal block is nothing but the unitary matrix $Q$ while the
upper left diagonal element is fixed such that $\det \underline{Q} = 1$. With
such a form of $\underline{Q}$, the action (\ref{Sext}) coincides with the
standard AIII class sigma-model action (\ref{action}) with a Wess-Zumino term
added. At the same time, the group $\mathrm{SU}(N + 1)$ is simply connected
hence the vortices are topologically trivial configurations in the extended
model. Inside a vortex core, massive elements of $\underline{Q}$ become non-zero
and provide an overall smooth field configuration. The size of the core is
determined by the competition between energy loss due to the mass $M$ and energy
gain due to avoiding large gradients. This yields the core size $\sim
\sqrt{\sigma}/M$.

Within the extended model we can examine the inner structure of the vortex
core. Assume for simplicity that the vortex occurs in the first replica, i.e.,
the corresponding unit vector is $|p\rangle = \{1, 0, \dots, 0\}$. The whole
vortex configuration, including the core, will involve only the upper-left $2
\times 2$ block of the matrix $\underline{Q}$. This block is an $\mathrm{SU}(2)$
matrix and we can explicitly parameterize it by three angles in the following
way:
\begin{equation}
 \underline{Q}
  = \begin{pmatrix}
      \cos\theta e^{-i\phi} & i \sin\theta e^{-i\chi} & 0\\
      i \sin\theta e^{i \chi} & \cos\theta e^{i\phi} & 0\\
      0 & 0 & \mathbbm{1}_{N-1 \times N-1}
    \end{pmatrix}.
 \label{uQvort}
\end{equation}
Far from the vortex core, the angle $\theta$ vanishes and the matrix
$\underline{Q}$ acquires the form (\ref{uQsoft}); it is independent of $\chi$.
Going around the vortex, angle $\phi$ rotates by $2\pi$. In the center of the
vortex $\theta = \pi/2$ and $\underline{Q}$ is independent of $\phi$ but
explicitly depends on $\chi$.  Using the ansatz (\ref{uQvort}), we can minimize
the action (\ref{Sext}). The symmetry of vortex allows us to fix parameter
$\phi$ equal to the polar angle and $\chi$ constant. We see that the vortex core
acquires an inner $\mathrm{U}(1)$ degree of freedom, $\chi$, which is beyond the
sigma model and is effective only in the extended theory. The action is
minimized by a proper $\theta(r)$ dependence. 

The Wess-Zumino term is responsible for the imaginary part of the action. We can
calculate this imaginary part without solving for $\theta(r)$ dependence. Let us
consider the variation of the Wess-Zumino term with respect to spatially
constant $\chi$. Upon substitution of Eq.\ (\ref{uQvort}) into Eq.\
(\ref{varWZ}), we obtain
\begin{multline}
 \delta S_\text{WZ}[\underline{Q}]
  = \frac{i k \delta\chi}{2\pi} \int d^2r\, \sin 2\theta \big(
      \nabla_x\theta \nabla_y\phi - \nabla_y\theta \nabla_x\phi
    \big) \\
  = ik \delta\chi \Big[ \sin^2\theta(r=\infty) - \sin^2\theta(r=0) \Big]
  = -ik \delta\chi.
 \label{dSdchi}
\end{multline}
Thus the value of the Wess-Zumino term explicitly depends on $\chi$.

The imaginary part of the action makes vortex fugacity complex. The arameter
$\chi$, that determines the phase of the complex fugacity, is an internal degree
of freedom of the vortex core. Calculating the partition function of the system,
we have to integrate over $\chi$ for each vortex. Once the action contains the
Wess-Zumino term, i.e., $k \neq 0$, such an integration will exactly cancel the
statistical weight of each vortex making fugacity effectively zero. This once
again demonstrates that the Wess-Zumino model of class AIII does not allow
vortex
excitations.

\subsection{Class CII with  $\mathbb{Z}_2$  $\theta$ term}
\label{sec:topologycii}

Let us now consider the system of symmetry class CII. The topology of the
target manifold admits the $\mathbb{Z}_2$ topological term. It is related to
the homotopy group $\pi_2(\mathrm{U}(N)/\mathrm{O}(N)) = \mathbb{Z}_2$. One
particular realization of the symmetry class CII is provided by the disordered
massless Dirac Hamiltonian of the form Eq.\ (\ref{H-chiral}) with
\begin{equation}
 h
  = \begin{pmatrix}
      p_x - i p_y & a \\
      -a^* & p_x + i p_y     
    \end{pmatrix}.
 \label{hCII}
\end{equation}
Here $a$ is a random $n \times n$ matrix with complex entries. The block $h$
obeys symplectic constraint $h^* = \tau_y h \tau_y$. In other words, $h$ is an
$n \times n$ matrix of real quaternions. Thus the Hamiltonian built from $h$
indeed belongs to the symmetry class CII.

The derivation of the sigma model for the disordered system described by the
Hamiltonian (\ref{H-chiral}), (\ref{hCII}) is outlined in Appendix
\ref{app:sigma}. The action of the model has the standard form (\ref{action})
with an additional $\mathbb{Z}_2$ topological term when $n$ is odd. This
topological term can be expressed in the form very similar to the Wess-Zumino
term (\ref{WZ}). Specifically, we have to continuously extend the $Q$ matrix to
the auxiliary third dimension $\tau$ according to Eq.\ (\ref{Qext}). Then the
topological term can be written in the form of Eq.\ (\ref{WZ}) with $k = n$.
Since the second homotopy group of the target manifold is non-trivial, the
extension (\ref{Qext}) is not always possible within the target space of class
CII. In fact, it is only possible for topologically trivial configurations of
$Q(\mathbf{r})$. To apply the Wess-Zumino construction to a general field
configuration, we will assume that away from the physical 2D space, at $\tau
\neq 1$, $Q$ is any unrestricted unitary matrix from $\mathrm{U}(N)$, while for
$\tau = 1$ it is unitary and symmetric hence belongs to the coset space
$\mathrm{U}(N)/\mathrm{O}(N)$ of the class CII.

Similarly to the class AIII, the value of the Wess-Zumino term is actually
determined by the physical part of $Q$ at $\tau = 1$ which is a unitary
symmetric matrix. The variation of the Wess-Zumino term with small variations of
$Q$ is given by Eq.\ (\ref{varWZ}). Using the property $Q = Q^T$ and transposing
the argument of the trace in the last line of Eq.\ (\ref{varWZ}), we see that
the variation changes sign and hence is identically zero. This proves that such
a Wess-Zumino term in the class CII possesses the main property of the theta
term: it depends only on the topology of the field configuration.

Apart from the class of topologically trivial configurations, there is only one
extra non-trivial class. In other words, the homotopy group $\pi_2 =
\mathbbm{Z}_2$ implies existence of localized topological ``excitations'',
$\mathbbm{Z}_2$ instantons. They are their own ``antiparticles''; configuration
of two such instantons is topologically trivial, i.e. the instantons can be
brought close to each other and annihilated by an appropriate continuous
transformation of $Q$. In order to prove that the Wess-Zumino term (\ref{WZ}),
being constant in each topological class, distinguishes between them, it
suffices to evaluate it for one particular non-trivial instanton configuration.

Let us consider the minimal model $Q \in \mathrm{U}(2)/\mathrm{O}(2)$. (In fact,
in this case the homotopy group $\pi_2 = \mathbb{Z}$ is reacher than in the
general case $N > 2$ and the theory possesses usual $\mathbb{Z}$ instantons
similar to, e.g., $\mathrm{O}(3)$ vector sigma model.) In fact, it is sufficient
to consider an even smaller target space $\mathrm{SU}(2)/\mathrm{O}(2)$ since
the determinant of $Q$ anyway drops from the Wess-Zumino term (\ref{WZ}). We
parametrize the $\mathrm{SU}(2)$ matrix by three angles in the following way:
\begin{equation}
 Q
  = \begin{pmatrix}
      \cos\theta \cos\chi e^{-i\phi} & i \sin\theta \cos\chi + \sin\chi \\
      i \sin\theta \cos\chi - \sin\chi & \cos\theta \cos\chi e^{i\phi}
    \end{pmatrix}.
 \label{SU2}
\end{equation}
The symmetry condition $Q = Q^T$ fixes $\chi = 0$ in the physical 2D space. For
the instanton, we can assume that $\phi$ is equal to the polar angle while
$\theta$ depends only on the radial coordinate and changes continuously from
$-\pi/2$ in the center of the instanton to $\pi/2$ at infinity. Extending to
the third dimension, we will assume $\chi$ to change from $0$ at $\tau = 1$ to
either $\pi/2$ or $-\pi/2$ at $\tau = 0$. Any of these two values uniquely fixes
the whole matrix $Q(\tau = 0)$.

Wess-Zumino action (\ref{WZ}) can be explicitly written in the parametrization
(\ref{SU2}) as
\begin{equation}
  S_\text{WZ}[Q]
  = \frac{i k}{\pi} \epsilon_{\mu\nu\lambda} \int d^2r\, d\tau 
      \cos\theta \cos^2\chi
      \nabla_\mu \theta \nabla_\nu \phi \nabla_\lambda \chi.
\end{equation}
For the instanton configuration, $\theta$ and $\chi$ depend on $r$ and $\tau$,
respectively, while $\phi$ is the polar angle. Calculating the integral, we
obtain
\begin{equation}
  S_\text{WZ}[Q]
  = \mp \frac{i \pi k}{2} \Big[ \sin\theta(r=\infty) - \sin\theta(r=0) \Big]
  = \mp i \pi k.
\end{equation}
The two signs in this expression correspond to the two extensions $\chi(\tau=0)
= \pm \pi/2$. For an odd value of $k$, the instanton action acquires a
non-trivial imaginary contribution. Thus the Wess-Zumino term indeed plays the
role of a theta term yielding $i \pi k$ times the topological charge of the
field configuration.

Once the explicit form of the $\mathbb{Z}_2$ topological term is established, we
can discuss its interplay with vortices. We have already argued that the
Wess-Zumino term in the sigma model of class AIII makes vortex excitations
ineffective. Similar arguments can be applied to the class CII with the
topological term since it has the same structure as the Wess-Zumino term. The
only difference in the class CII is an additional constraint related to the
time-reversal symmetry. Namely, the statistical weight of any field
configuration of the class CII sigma model must be real. Equivalently, the
imaginary part of the sigma-model action must be an integer multiple of $i\pi$.
It is the topological term that provides this imaginary part. Consider an
extension of the model from $\mathrm{U}(N)/\mathrm{O}(N)$ up to
$\mathrm{SU}(N+1)/\mathrm{O}(N+1)$. Such an extension is given by, e.g., Eq.\
(\ref{Sext}) with an additional symmetry constraint $\underline{Q} =
\underline{Q}^T$. The vortex configuration in this extended model has the form
Eq.\ (\ref{uQvort}) with the angle $\chi$ taking either $0$ or $\pi$ value. Thus
the internal $\mathrm{U}(1)$ parameter associated with the vortex in the
symmetry class AIII becomes a $\mathbb{Z}_2$ degree of freedom in the class CII.
The two vortex configurations with $\chi = 0$ and $\chi = \pi$ differ by an
$i\pi$ term in the action, cf.\ Eq.\ (\ref{dSdchi}). Thus summation over the two
values of this internal degree of freedom effectively annihilates vortex
contribution to the partition function of the system.

The above discussion is based on the specific form  of the extended model
(\ref{Sext}) describing the vortex core. In fact, we can lift this restriction
and show that a vortex possesses an internal $\mathbb{Z}_2$ degree of freedom
without assuming any particular structure of its core. Consider a vortex
configuration in the class CII sigma model. We suppose that some additional
massive degrees of freedom become relevant in the center of the vortex and once
they are taken into account the field configuration is continuous. The extended
model must possess the time-reversal symmetry characteristic for the class CII.
Hence the statistical weight in the extended model is real and the action is
real up to an integer multiple of $i\pi$. Let us now create a small instanton
far from the vortex center. This will change the imaginary part of the action by
$\pi$. Within the extended theory, we can bring the instanton close to the
vortex center and ``hide'' it inside the core by a continuous field
transformation. Since the imaginary part of the action takes only discrete
values, this transformation will not remove an extra $i\pi$ from the action
related to the instanton. We have thus demonstrated the existence of two
topologically distinct vortex solutions in the extended theory with opposite
signs of their statistical weights. Since there is no other general distinction
between these two solutions, the real parts of their action must be equal. This
once again shows that the total statistical weight of a vortex configuration is
zero if the underlying sigma model contains the $\mathbb{Z}_2$ topological
term. 

To conclude, additional topological terms in the sigma model of both AIII and
CII symmetry classes suppress formation of vortices and thus prevent the system
from localization.

\section{Summary and outlook}
\label{sec:summary}

In this paper, we have developed a field-theoretical (sigma-model) approach to
Anderson localization in 2D disordered systems of chiral symmetry classes (AIII,
BDI, CII). A remarkable feature of sigma models for these classes is that the
quantum interference effects leading to renormalization of conductivity (and
thus to Anderson localization) are absent to all orders of perturbation theory.
We have shown that Anderson localization does exist within these models and is
governed by a non-perturbative mechanism. Specifically, the localization is due
to topological excitations -- vortices -- of the sigma model field. We have
derived the corresponding renormalization group equations which include
non-perturbative contributions. Analyzing them, we find that the 2D disordered
systems of chiral classes undergo a metal-insulator transition driven by
topologically induced Anderson localization.

While the mechanism of the localization transition --- proliferation of vortices
--- bears an analogy with the Berezinskii-Kosterlitz-Thouless transition in
systems with $\mathrm{U}(1)$ symmetry, our RG equations are essentially
different. The reason for this is a more complex structure of the theory: it is
characterized by three coupling constants (conductivity $\sigma$, Gade coupling
$c$, and fugacity $y$) instead of two couplings of the BKT transition theory
(spin stiffness and fugacity). As a result, the fixed point governing the
transition turns out to be at non-zero fugacity. For this reason, the critical
behavior at the transition cannot be determined in a controllable way. The
one-loop analysis suggests that this behavior is of power-law type (i.e. is more
similar to the critical behavior at Anderson transitions in conventional classes
rather than to that at BKT transition). 

For the chiral unitary class AIII, we have presented an alternative derivation
of the renormalization group based on a mapping of the sigma model onto a dual
theory. The latter has the form of a generalized sine-Gordon theory. 

We have also considered 2D disordered systems formed on surfaces of 3D
topological insulators of chiral symmetry classes AIII and CII. In this case the
sigma model is supplemented by a term of topological origin: the Wess-Zumino
term for the class AIII and $\mathbb{Z}_2$ theta term for the class CII. We have
shown that such terms overpower the effect of vortices, thus ensuring the
protection of surface states against the vortex-induced Anderson localization.

Our work opens perspectives for research in a number of important directions.
Below, we briefly discuss several of them.

First, it would be very interesting to investigate the metal-insulator
transition in chiral classes and the associated critical behavior numerically.
Remarkably, this issue is almost unexplored by now. This is in stark contrast
with conventional symmetry classes (i.e., symplectic class metal-insulator
transition and unitary class quantum Hall transition in 2D, orthogonal class
transition in 3D etc.) where very detailed studies have been carried out. Since 
we are dealing here with non-interacting systems in a relatively low (2D)
dimensionality, a sufficiently accurate numerical analysis is expected to be
feasible. It would be also interesting to simulate the sigma model directly.
This would allow to verify the importance of topological excitations for
localization and to test our predictions. Such an approach can be implemented
within supersymmetric version of the sigma model with Grassmann
degrees of freedom integrated out.\cite{Disertori}

Second, it remains to develop a dual, sine-Gordon-like theory of the
transition, analogous to that presented in Sec.\ {\ref{sec:sine-gordon} for
class AIII, for the other two chiral classes. Furthermore, we feel that
geometric aspects of sigma-model renormalization within this dual formalism
deserve a more thorough investigation. 

Third, a natural question arises concerning the metal-insulator transitions in
chiral classes in 3D (and higher dimensionalities). We expect that also there
the transition will be driven by topological excitations, namely, vortex lines.

Fourth, in analogy with $\mathbb{Z}$ vortices studied above, 2D sigma models of
two classes, AII and DIII, allow for $\mathbb{Z}_2$ vortices. The difference is
that these two classes do show quantum interference effects on the perturbative
level. Therefore, in contrast to the chiral classes, the vortices in classes AII
and DIII will not constitute the only driving mechanism of Anderson localization
but rather will contribute to renormalization of conductivity along with
perturbative terms.

Fifth, interaction effects play an important role in low-dimensional systems
and may strongly affect the nature of the metal-insulator transition.
Two-dimensional Dirac fermions subjected to short-range interaction exhibit the
Mott transition\cite{FosterLudwig} unlike the non-interacting case when the
system remains metallic. It would be very interesting to investigate the
interplay of interaction and vortices in systems with chiral symmetry.

Finally, we close with a more general comment. The importance of topological
aspects of field theories of disordered systems was recently emphasized in the
context of topological insulators \cite{SchnyderKitaev, AL50}. There, a
possibility of emergence of a Wess-Zumino term or $\mathbb{Z}_2$ theta term in
the corresponding dimensionality and symmetry class signals the existence of a
topological insulator phase. There is at present a growing appreciation of the
fact that the topological properties of sigma-model manifolds may crucially
affect the physical observables even for those combinations of dimensionalities
and symmetries than do not allow for topological insulators. In particular, a
recent work \cite{Gruzberg11} has shown that a particular symmetry of local
density of states distributions holds for sigma models (and thus for critical
points) of five symmetry classes (A, AI, AII, C, CI) but dos not hold for the
remaining five (AIII, BDI, CII, D, DIII). The distinct feature of the latter
five classes is the presence of $\mathrm{U}(1)$ or $\mathrm{O}(1)$ subgroup in
the sigma model target space $\mathcal{M}$ leading to a non-trivial topology:
$\pi_1(\mathcal{M}) = \mathbb{Z}$ for chiral classes (AIII, BDI, CII) and 
$\pi_0(\mathcal{M}) = \mathbb{Z}_2$ for Bogoliubov-de Gennes classes D and DIII.
It is worth emphasizing that these topologies render these five classes
topological insulators in 1D. Equivalently, models of these symmetry classes may
support eigenstates with exactly zero energies \cite{Ivanov02}. As the paper
Ref.~\onlinecite{Gruzberg11} showed, the same $\mathrm{U}(1)$ and
$\mathrm{O}(1)$ degrees of freedom that are responsible for topological
insulator properties in 1D in fact crucially affect the multifractal spectra at
higher dimensionalities. The present work shows that the $\mathrm{U}(1)$
topology of sigma-model manifolds of chiral classes is also responsible for
Anderson localization in these classes in 2D (and likely also in higher
dimensions). Earlier, the authors of Ref.\ \onlinecite{Bocquet00} argued that
the $\mathrm{O}(1) = \mathbb{Z}_2$ degree of freedom of the sigma model is
responsible for localization in class D in two dimensions. Thus, it turns out
that the importance of topological aspects of field theories of disordered
systems goes well beyond that expected on the basis of classification of
topological insulators. Full ramifications of these observations remain to be
understood.

\section{Acknowledgments}
\label{sec:acknowledgments}

We are grateful to I.\ V.\ Gornyi, I.\ A.\ Gruzberg, V.\ Gurarie, D.\ A.\
Ivanov, A.\ W.\ W.\ Ludwig, and M.\ A.\ Skvortsov for valuable discussions. The
work was supported by the Center for Functional Nanostructures and the SPP 1459
``Graphene'' of the Deutsche Forschungsgemeinschaft and by the German Ministry
for Education and Research (BMBF). The work of I.\ V.\ P.\ was supported by
Alexander von Humboldt Foundation. P.\ M.\ O.\ and A.\ D.\ M.\ acknowledge the
hospitality of KITP Santa Barbara at the stage of preparation of the manuscript
for publication.

\appendix

\section{Perturbative renormalization in dual representation}
\label{app:theta_renormalization}

The purpose of this appendix is to illustrate  the equivalence of the model
(\ref{Z_theta}), (\ref{S_theta}) to the original AIII sigma model by a one-loop
perturbative renormalization. Within this calculation we can ignore the Gaussian
field $\theta_0$ in Eq.\ (\ref{S_theta}) (since renormalization of this sector
is trivial on the perturbative level) and concentrate on the renormalization of
the non-abelian sector of the theory parametrized by fields $\theta_a$ with $a =
1, \ldots N^2 - 1$. To make more explicit the possibility of a loop expansion
controlled by the parameter $\sigma \gg 1$, it is also convenient to rescale
$\theta_a$ by a factor $\sigma/4\pi$. The action of the model acquires now the
form 
\begin{gather}
 S[\theta]
  = \frac{\sigma}{8\pi} \big( \delta_{\mu\nu} + i\epsilon_{\mu\nu} \big)
   \int d^2r\, (\nabla_\mu\theta_a) g_{ab}(\theta) (\nabla_\nu\theta_b), \\
 g(\theta)
  = \big[ \mathbbm{1} - Z(\theta) \big]^{-1},
 \qquad
 Z_{ab}(\theta)
  = \sum_{c = 1}^{N^2 - 1} \theta_c f_{abc}.  
\end{gather}

The renormalization of the theory can now be carried out in the standard manner.
We split the fields $\theta^a$ into the fast and slow components, $\theta =
\theta_< + \tilde\theta_>$. We should then expand in the fast fields up to the
second order and integrate them out. It proves convenient to perform the change
of fast variables $\theta_> = g(\theta_<) \tilde\theta_>$ so that the
decomposition of $\theta$ into the fast and slow fields reads 
\begin{equation}
 \theta
  = \theta_< + \big( 1 - Z_< \big) \theta_>.
\end{equation}
Here we introduced for notational brevity $Z_< \equiv Z(\theta_<)$. This change
of integration variables cancels the contribution of the non-trivial integration
measure $D\mu(\theta)$ to the renormalization group equations  and guarantees
the absence of linearly diverging diagrams in one-loop calculation. Denoting
also $Z(\theta_>)$ by $Z_>$ we have 
\begin{align}
 \nabla\theta
  &= (1 + Z_>) \nabla \theta_< + (1 - Z_<) \nabla\theta_>, \\
 Z(\theta)
  &= Z_< + Z_> + [Z_>, Z_<],
\end{align}
This implies that the quadratic-in-$\theta_>$ contribution to the action reads
$S_2 = S_2^{(0)} + \delta S_2$, 
\begin{widetext}
\begin{align}
 S_2^{(0)}
  &= \frac{\sigma}{8\pi} \int d^2r\, \nabla\theta_>^T \nabla\theta_>, \\
 \delta S_2
  &= \frac{\sigma (\delta_{\mu\nu} + i\epsilon_{\mu\nu})}{8\pi}
    \int d^2r\, \big[
      \nabla_\mu \theta_>^T (1 + Z_<)
      +\nabla_\mu \theta_<^T g_< Z_> Z_<
    \big]
    \big[\nabla_\nu \theta_> + g_< (2 - Z_<) Z_> g_< \nabla_\nu \theta_<\big]
    -S_2^{(0)}.
\end{align}
\end{widetext}

Performing now the expansion in $\delta S_2$, averaging over fast fluctuations,
and retaining only the logarithmically diverging contributions, we find the
correction to the action functional of the slow fields: 
\begin{multline}
 \Delta S
  = \frac{N}{8\pi} \ln\frac{\Lambda}{\Lambda'}
    (\delta_{\mu\nu} + i\epsilon_{\mu\nu}) \\
  \times \int  d^2r\,
    \nabla_\mu\theta^T_< \big[2 g + 3 g^2 \big]\nabla_\nu \theta_<\,. 
\end{multline}

We can recast the action for the slow fields into the original form by
correcting the conductivity $\sigma \mapsto \tilde\sigma$ and switching to the
rescaled filed $\tilde\theta$ 
\begin{align}
 \tilde{\sigma}
  &= \sigma \left(
      1 - \frac{N}{\sigma} \ln\frac{\Lambda}{\Lambda'}
    \right), \\ 
 \tilde{\theta}
  &= \theta \left(
      1 + \frac{3N}{\sigma} \ln\frac{\Lambda}{\Lambda'}
    \right).
\end{align}
(Note that, after the rescaling, the factor $1/(1-Z)$ should be expanded to the
first order in $1/\sigma$, which is the accuracy of the one-loop calculation.)
We see now that the dual model defined by Eqs.\ (\ref{Z_theta}), (\ref{S_theta})
reproduces the correct renormalization of the conductivity $\sigma$.

\section{Chiral sigma model for massless Dirac fermions}
\label{app:sigma}

In this Appendix we outline the derivation of the sigma models for the massless
Dirac Hamiltonians of AIII and CII symmetries. Apart from demonstrating the
consequences of the chiral symmetry in the sigma-model language, we will also
discuss possible topological terms. Such terms frequently appear due to chiral
anomaly of the massless Dirac electrons.

\subsection{Class AIII}

Let us first consider the AIII symmetry class. A general chiral Hamiltonian has
the block off-diagonal structure (\ref{H-chiral}). We consider the following
Dirac Hamiltonian of AIII symmetry [cf.\ Eq.\ (\ref{HAIII})]:
\begin{equation}
 H
  = \begin{pmatrix}
      0 & p_- + a \\
      p_+ + a^\dagger & 0
    \end{pmatrix}.
\end{equation}
Here $p_\pm = p_x \pm i p_y$ and $a$ is a random complex-valued matrix of size
$n \times n$. Further, we adopt the most standard model of Gaussian white noise
disorder with the correlator
\begin{equation}
 \langle a^*_{ij} a_{kl} \rangle
  = \frac{\alpha}{n}\; \delta_{ik} \delta_{jl}.
 \label{Gauss}
\end{equation}
Parameter $\alpha$ quantifies the disorder strength and determines the
scattering rate of electrons. Within self-consistent Born approximation, the
scattering rate at the Dirac point (zero chemical potential) is given by
\begin{equation}
 \gamma
  = \Delta\, e^{-2\pi/\alpha},
 \label{gamma}
\end{equation}
where $\Delta$ is the effective band width (maximal allowed energy) for Dirac
Hamiltonian. The self-consistent Born approximation is valid in the limit $n \gg
1$. This also corresponds to high Drude conductivity $n/\pi (e^2/h)$ at the
Dirac point and justifies the applicability of the sigma model.

Derivation of the non-linear sigma model starts with the replicated action
written in terms of fermionic (anticommuting) fields:
\begin{equation}
 S
  = \begin{pmatrix} \bar\psi_i^a & \bar\phi_i^a \end{pmatrix}
    \begin{pmatrix}
      0 & \delta_{ij} p_- + a_{ij} \\
      \delta_{ij} p_+ + a^*_{ji} & 0
    \end{pmatrix}
    \begin{pmatrix} \phi_j^a \\ \psi_j^a \end{pmatrix}.
 \label{Spsihpsi}
\end{equation}
The lower indices $i$ and $j$ take $n$ possible values in the flavor space while
the upper index $a$ enumerate $N$ replicas. We proceed with averaging $e^{-S}$
over the random matrix $a$. With the help of Eq.\ (\ref{Gauss}), the effective
action acquires the form
\begin{gather}
 S
  = \bar\psi_i^a p_- \psi_i^a + \bar\phi_i^a p_+ \phi_j^a
    +S_\text{dis}, \label{SpsiAIII}\\
 S_\text{dis}
  = \frac{\alpha}{n}\; \psi_i^a \bar\phi_i^b \phi_j^b \bar\psi_j^a.
 \label{SdisAIII}
\end{gather}
The quartic term $S_\text{dis}$ is further decoupled with the help of an
auxiliary complex matrix field $Q$ acting in replica space. This is achieved by
adding to the action a term $\mathop{\mathrm{Tr}} Q^\dagger Q$ and shifting the
$Q$ variable by a suitable quadratic expression in fermions.
\begin{multline}
 S_\text{dis}
  \mapsto S_\text{dis} + \frac{n \gamma^2}{\alpha}
    \left[Q_{ab} + \frac{i \alpha}{n \gamma} \psi_i^a \bar\phi_i^b \right]
    \left[Q^*_{ab} + \frac{i \alpha}{n \gamma} \phi_j^b \bar\psi_j^a \right] \\
  = \frac{n \gamma^2}{\alpha} \mathop{\mathrm{Tr}} Q^\dagger Q
    -i \gamma \Big(
      \bar\psi_i^a  Q_{ab} \phi_i^b + \bar\phi_i^b Q^*_{ab} \psi_i^a
    \Big).
\end{multline}
Now we add the rest of the action from Eq.\ (\ref{SpsiAIII}) and then perform
Gaussian integration over fermion fields.
\begin{multline}
 S
  = \frac{n \gamma^2}{\alpha} \mathop{\mathrm{Tr}} Q^\dagger Q
    +\begin{pmatrix} \bar\psi_i & \bar\phi_i \end{pmatrix}
    \begin{pmatrix} -i\gamma Q & p_- \\ p_+ & -i\gamma Q^\dagger \end{pmatrix}
    \begin{pmatrix} \phi_i \\ \psi_i \end{pmatrix} \\
  \mapsto \frac{n \gamma^2}{\alpha} \mathop{\mathrm{Tr}} Q^\dagger Q
    - n \mathop{\mathrm{Tr}} \ln
    \begin{pmatrix} -i\gamma Q & p_- \\ p_+ & -i\gamma Q^\dagger \end{pmatrix}.
 \label{SQAIII}
\end{multline}

The next step of the sigma model derivation involves saddle-point analysis of
the above action for $Q$. The saddle-point equation is equivalent to the
equation of self-consistent Born approximation with $i\gamma Q$ playing the role
of self energy. Therefore one particular solution is just the unit matrix $Q =
\mathbbm{1}$. Other solutions can be found by unitary rotating the fermion
fields in the first line of Eq.\ (\ref{SQAIII}) such that the $p_\pm$ terms
remain intact. These rotations generate the gauge group $\mathrm{U}(N) \times
\mathrm{U}(N)$ rotating $Q$ by the two unitary matrices from left and right.
Thus the matrix $Q$ takes values from the symmetric subgroup $\mathrm{U}(N)
\times \mathrm{U}(N) / \mathrm{U}(N) = \mathrm{U}(N)$ of the global gauge group.
This establishes the manifold of the class AIII sigma model.

The sigma-model action is a result of the gradient expansion of Eq.\
(\ref{SQAIII}) with a slowly varying unitary matrix $Q$. This gradient expansion
should be carried out with care in view of the chiral anomaly of the Dirac
operator under logarithm. A systematic description of the expansion procedure,
including the methods to treat the anomaly, can be found, in particular, in
Ref.\ \onlinecite{ASZ}. The result of the gradient expansion reads
\begin{equation}
 S[Q]
  = -\frac{n}{8\pi} \mathop{\mathrm{Tr}} \big(Q^\dagger \nabla Q \big)^2
    +S_\text{WZ}[Q].
 \label{SAIIIWZ}
\end{equation}
This is the standard sigma-model action of class AIII, Eq.\ (\ref{action}) with
$\sigma = n$ and $c = 0$, with an additional Wess-Zumino term of the level $k =
n$, see Eq.\ (\ref{WZ}). Appearance of the Wess-Zumino term is the direct
consequence of the chiral anomaly of the Dirac Hamiltonian. The structure and
properties of the Wess-Zumino term are discussed in the main text, Sec.\
\ref{sec:topologyaiii}.

\subsection{Class CII}

Now we discuss the derivation of the sigma model with $\mathbb{Z}_2$ topological
term in the symmetry class CII. Consider the Hamiltonian [cf.\ Eq.\
(\ref{hCII})]
\begin{equation}
 H
  = \begin{pmatrix}
      0 & h \\
      h^\dagger & 0
    \end{pmatrix},
 \qquad
 h
  = \begin{pmatrix}
      p_- & a \\
      -a^* & p_+
    \end{pmatrix}.
 \label{CII}
\end{equation}
As in the previous case, $a$ is a random complex-valued matrix of size $n$. 
The block $h$ fulfills the symmetry condition $h = \tau_y h^* \tau_y$ and hence
represents a (operator-valued) real quaternion matrix. Thus the Hamiltonian
(\ref{CII}) indeed belongs to the symmetry class CII.

We assume the same Gaussian white-noise distribution of $a$ as in the previous
section, Eq.\ (\ref{Gauss}). Disorder-induced scattering rate, Eq.\
(\ref{gamma}), is also reproduced within the self-consistent Born approximation;
Drude conductivity is twice larger, $2n/\pi (e^2/h)$, since the Hamiltonian
(\ref{CII}) contains $2n$ coupled Dirac fermions.

As in the previous section, we start with the replicated fermionic action,
\begin{multline}
 S
  = \begin{pmatrix} \bar\psi & \bar\xi \end{pmatrix}_i^a
    h_{ij}
    \begin{pmatrix} \psi \\ \xi \end{pmatrix}_j^a
    +\begin{pmatrix} \bar\phi & \bar\zeta \end{pmatrix}_i^a
    h_{ji}^*
    \begin{pmatrix} \phi \\ \zeta \end{pmatrix}_j^a \\
  = \bar\Psi^a_i p_- \Psi^a_i + \bar\Phi^a_i p_+ \Phi^a_i
    +\bar\Psi^a_i a_{ij} \bar\Phi^a_j - \Phi^a_i a^*_{ij} \Psi^a_j.
 \label{SpsiCIIbare}
\end{multline}
Here we have introduced the doubled fermionic fields
\begin{equation}
 \Psi
  = \begin{pmatrix} \psi \\ \bar\zeta \end{pmatrix}, \quad
 \Phi
  = \begin{pmatrix} \bar\xi \\ \phi \end{pmatrix}, \quad
 \bar\Psi
  = \begin{pmatrix} \bar\psi \\ \zeta \end{pmatrix}, \quad
 \bar\Phi
  = \begin{pmatrix} \xi \\ \bar\phi \end{pmatrix}.
 \label{Psi}
\end{equation}
Starting from the second line of Eq.\ (\ref{SpsiCIIbare}) and further on, we
assume that the replica index $a$ takes on $2N$ values running through $N$
replicas and two components of the vectors (\ref{Psi}).

We average $e^{-S}$ from Eq.\ (\ref{SpsiCIIbare}) using the correlator Eq.\
(\ref{Gauss}) and obtain
\begin{gather}
 S
  = \bar\Psi^a_i p_- \Psi^a_i + \bar\Phi^a_i p_+ \Phi^a_i
    +S_\text{dis}, \label{SpsiCII}\\
 S_\text{dis}
  = \frac{\alpha}{n}\; \Psi_i^b \bar\Phi_i^a \Phi_j^b \bar\Psi_j^a.
 \label{SdisCII}
\end{gather}
These expression are very similar to Eqs.\ (\ref{SpsiAIII}) and
(\ref{SdisAIII}). The only difference is in the order of replica indices in the
quartic term Eq.\ (\ref{SdisCII}). As before, we introduce the matrix field $Q$
and decouple the action according to
\begin{multline}
 S_\text{dis}
  \mapsto S_\text{dis} + \frac{n \gamma^2}{\alpha}
    \left[Q_{ab} + \frac{i \alpha}{n \gamma} \Psi_i^b \bar\Phi_i^a \right]
    \left[Q^*_{ab} + \frac{i \alpha}{n \gamma} \Phi_j^b \bar\Psi_j^a \right] \\
  = \frac{n \gamma^2}{\alpha} \mathop{\mathrm{Tr}} Q^\dagger Q
    -i \gamma \Big(
      \bar\Psi_j^a  Q_{ab} \Phi_j^b + \bar\Phi_i^a Q^*_{ab} \Psi_i^b
    \Big).
\end{multline}
Adding the clean part of the action from Eq.\ (\ref{SpsiCII}) and performing
Gaussian integration over fermion fields, we obtain the action in the form
\begin{multline}
 S
  = \frac{n \gamma^2}{\alpha} \mathop{\mathrm{Tr}} Q^\dagger Q
    +\begin{pmatrix} \bar\Psi_i & \bar\Phi_i \end{pmatrix}
    \begin{pmatrix} -i\gamma Q & p_- \\ p_+ & -i\gamma Q^* \end{pmatrix}
    \begin{pmatrix} \Phi_i \\ \Psi_i \end{pmatrix} \\
  \mapsto \frac{n \gamma^2}{\alpha} \mathop{\mathrm{Tr}} Q^\dagger Q
    - n \mathop{\mathrm{Tr}} \ln
    \begin{pmatrix} -i\gamma Q & p_- \\ p_+ & -i\gamma Q^* \end{pmatrix}.
 \label{SQCII}
\end{multline}
This result is again very similar with the class AIII expression
(\ref{SQAIII}). The only difference is that the matrix $Q^\dagger$ under the
logarithm in Eq.\ (\ref{SQAIII}) is replaced with $Q^*$ in the present case.
The standard saddle point of the above action is $Q = \mathbbm{1}$. Other saddle
points are generated by rotations of the fermion fields in the first line of
Eq.\ (\ref{SQCII}). As follows from the saddle point analysis, these rotations
should maintain the complex conjugacy of $Q$ and $Q^*$ in the argument of the
logarithm. This is achieved by the unitary transformation of the form $\Phi
\mapsto U \Phi$, $\Psi \mapsto U^* \Psi$, $\bar\Phi \mapsto \bar\Phi U^\dagger$,
$\bar\Psi \mapsto \bar\Psi U^T$ parametrized by the unitary matrix $U$ of the
size $2N$. The global gauge group is thus $\mathrm{U}(2N)$. The saddle manifold
is parametrized as $Q = U^T U$ and contains all symmetric unitary matrices, $Q
\in \mathrm{U}(2N)/\mathrm{O}(2N)$, as it should be for the sigma model of class
CII.

For any symmetric unitary matrix $Q$, the action (\ref{SQCII}) is identical to
Eq.\ (\ref{SQAIII}). This means that the sigma-model action Eq.\
(\ref{SAIIIWZ}) obtained by the gradient expansion of Eq.\ (\ref{SQAIII}), is
also valid for the symmetry class CII provided the constraint $Q = Q^T$ is
maintained. With this restriction, the Wess-Zumino term of the level $k = n$
becomes a $\mathbb{Z}_2$ topological term for any odd $n$, as we discuss in
Sec.\ \ref{sec:topologycii}.\cite{Ryu11}

\end{document}